\documentclass[preprint]{elsarticle}

\usepackage{lineno,hyperref}
\modulolinenumbers[5]
\usepackage[cmex10]{amsmath}
\usepackage{amssymb}
\usepackage{subfigure}
\usepackage[flushleft]{threeparttable}
\usepackage{color, soul}
\journal{Journal of Medical Image Analysis}

\begin{document}

\begin{frontmatter}



%
%

\title{Deeply-Supervised Density Regression for Automatic Cell Counting in Microscopy Images}

\author{Shenghua He$^\text{a}$, Kyaw Thu Minn$^\text{b,c}$, Lilianna Solnica-Krezel$^\text{c,d}$, Mark A. Anastasio$^\text{e,*}$, and Hua Li$^\text{e,f,g,*}$}
\address{
$^\text{a}$Department of Computer Science and Engineering,\break Washington University in St. \!Louis, St. \!Louis, MO 63110 USA \\ 
$^\text{b}$Department of Biomedical Engineering,\break  Washington University in St. \!Louis, St. \!Louis, MO 63110 USA \\ 
$^\text{c}$Department of Developmental Biology,\break  Washington University School of Medicine in St. \!Louis, St. \!Louis, MO 63110 USA \\ 
$^\text{d}$Center of Regenerative Medicine,\break Washington University School of Medicine in St. \!Louis, St. \!Louis, MO 63110 USA \\ 
$^\text{e}$Department of Bioengineering,\break  University of Illinois at Urbana-Champaign, Urbana, IL 61801, USA \\
$^\text{f}$Cancer Center at Illinois,\break University of Illinois at Urbana-Champaign, Urbana, IL 61801 USA \\
$^\text{g}$Carle Cancer Center, Carle Foundation Hospital, Urbana, IL 61801 USA \\
$^\text{*}$Corresponding Author \\
}

\begin{abstract}
Accurately counting the number of cells in microscopy images is required in many medical diagnosis and biological studies. This task is tedious, time-consuming, and prone to subjective errors.
However, designing automatic counting methods remains challenging due to low image contrast, complex background, large variance in cell shapes and counts, and significant cell occlusions in two-dimensional microscopy images.
In this study, we proposed a new density regression-based method for automatically counting cells in microscopy images.
The proposed method processes two innovations compared to other state-of-the-art density regression-based methods.
First, the density regression model (DRM) is designed as a concatenated fully convolutional regression network (C-FCRN) to employ multi-scale image features for the estimation of cell density maps from given images.
Second, auxiliary convolutional neural networks (AuxCNNs) are employed to assist in the training of intermediate layers of the designed C-FCRN to improve the DRM performance on unseen datasets.
Experimental studies evaluated on four datasets demonstrate the superior performance of the proposed method.
\end{abstract}


\vspace{-0.2in}
\end{frontmatter}


\vspace{-0.2in}
\section{Introduction}
\label{sec:intro}

Numerous microscopy image analysis methods have been proposed for various medical diagnoses and biological studies that include counting the number of cells~\cite{lempitsky2010,arteta2014,xie2018}, locating cell positions~\cite{xing2016robust,falk2019u,koyuncu2020deepdistance}, acquiring cell shapes~\cite{zhang2016large, wainberg2018deep, xing2017deep,carneiro2017review}, and classifying cell categories~\cite{irshad2013methods,chen2016deep, hu2020label}.
Especially, the number of cells in a microscopy image can indicate the presence of diseases~\cite{venkatalakshmi2013automatic}, help differentiate tumor types~\cite{coates2015tailoring},
assist in understanding cellular and molecular genetic mechanisms~\cite{solnica2005conserved,zhang2001vitro},
and provide useful information to many other applications~\cite{thomson1998embryonic,lagutin2003six3}.
Manually counting cells in microscopy images is tedious, time-consuming, prone to subjective errors, and not feasible for high-throughput process in real-world biomedical applications.
During the past decades, many automatic cell counting methods have been proposed~\cite{matas2004robust,barinova2012,arteta2012,xing2014automatic,arteta2016,cirecsan2013,he2019automatic2}.
However, designing efficient automatic methods with sufficient counting accuracy still remains a challenging task
due to various image acquisition techniques, low image contrast, complex tissue background,
large variations in cell sizes, shapes and counts, and significant inter-cell occlusions in two-dimensional (2D) microscopy images.

The reported automatic cell counting methods can be categorized into \textit{detection-based} and \textit{regression-based} methods.
Generally, detection-based methods first determine the cell centroid locations and subsequently count them to estimate the number of cells~\cite{arteta2016,cirecsan2013,xing2014automatic,liu2017}.
Therefore, the performance of these methods highly relies on the accuracy of cell centroid detection results.
Traditional detection-based methods have been designed based on feature extraction~\cite{sommer2012learning}, morphological processing~\cite{soille2013morphological},
H-minima/maxima transform~\cite{soille2013morphological},
Laplacian of Gaussian filtering~\cite{kong2013generalized},
maximally stable extremal region detection~\cite{arteta2016},
radial symmetry-based voting~\cite{reisfeld1995context},
or conventional supervised learning strategies~\cite{xing2016robust}.
Recently, deep learning strategies have shown superior ability of extracting informative image features and generating inferences in all kinds of medical image analysis tasks~\cite{he2018convolutional,cirecsan2013,he2020learning}. A bunch of deep learning-based detection methods have been proposed~\cite{liu2017,zhu2017extended,tofighi2019prior,carneiro2017review,xing2017deep,xie2018efficient,bischof2019segmenting, koyuncu2020deepdistance, falk2019u}. For example, Falk et al.~\cite{falk2019u} trained a fully convolutional neural network (U-Net)
to compute a probability map of cell existing in a given image.
The number of cells can then be determined by searching for the local maxima on the probability map with a non-maxima suppression method.
Xie et al.~\cite{xie2018efficient} applied the non-maxima suppression process to a dense proximity map for cell detection.
The proximity map was produced by a fully residual convolutional network-based structural regression model (StructRegNet), and exhibits higher responses at locations near cell centroids to benefit for local maximum searching.
Tofighi et al.~\cite{tofighi2019prior} used {\it{a prior}}-guided deep neural network for cell nuclei detection.
In their method, nuclei shape {\it{a prior}} is employed as a regularizer in a model learning process to improve the cell detection accuracy.
Liu et al.~\cite{liu2017} trained a CNN model to determine the final cell detection result from the results generated by several traditional cell counting methods.
The selection process was formulated as a maximum-weight independent set (MWIS) problem,
a combinatorial optimization problem that has been studied in many applications of clustering, segmentation, and tracking.
Paulauskaite et al.~\cite{paulauskaite2019deep} recently performed an experimental investigation of the Mask R-CNN method, which was proposed by He et al.~\cite{he2017mask}, to detect overlapping cells with a two-stage procedure of determining potential cell regions and jointly classifying and predicating cell masks.
The method was validated on fluorescence and histology images and showed promising results on detecting overlapping cells.
However, it still remains difficult to detect cells that are highly occluded, densely concentrated, and surrounded by histopathological structures.

Compared to detection-based methods, regression-based cell counting methods have received more and more attention due to their superior performance on counting occluded cells~\cite{khan2016,xue2016,cohen2017,lempitsky2010, fiaschi2012, arteta2014,walach2016, xie2018,nitta2018intelligent,alahmari2019automated}
Some regression-based methods learn a cell counter through a regression process directly without requiring cell detection.
In these methods, the number of cells is the direct and only output, and no cell location information can be provided.
For example, Khan et al.~\cite{khan2016} and Xue et al.~\cite{xue2016} learned a convolutional neural network-based cell counter from small image patches which can increase the amount of training samples.
The total number of cells across the whole image can then be obtained by summing those on image patches. 
These methods might suffer from redundant estimation issues across the patch boundaries, and might not be efficient since they have to infer for each image patch separately before cell counting.
Differently, Cohen et al.~\cite{cohen2017} learned a cell counter with a fully convolutional neural network (FCNN).
They utilized the ``sliding window'' mechanism associated with the convolutional layers of the FCNN
to address the redundant counting issues across the overlapped regions among image patches. Their method counts the number of cells by directly inferring a count map for the whole image.
The method performance might be affected by the sizes of sliding widows.

Other regression-based methods learn a spatial cell density regression model (DRM)
across a full-size image instead of learning direct cell counters~\cite{lempitsky2010,fiaschi2012,xie2018,liu2019automated}.
In these methods, the number of cells can be obtained by integrating the regressed density map,
and the local maxima in the density map can be considered as cell centroid locations.
Therefore, both the number and the centroid locations of cells can be obtained.
Conventional density regression-based methods learn DRMs from extracted handcrafted image features, in which the feature extraction is independent of the DRM learning.
For example, Lempitsky et al.~\cite{lempitsky2010} used local features (e.g. scale-invariant feature transform (SIFT) features) to learn a linear DRM by use of a regularized risk regression-based learning framework.
Differently, Fiaschi et al.~\cite{fiaschi2012} learned a nonlinear DRM based on regression random forest methods.
In their method, image features computed by ordinary filter banks were employed as the model input.
The performance of these methods relies on the effectiveness of feature extraction methods, that of the DRM learning algorithms, and the match between them.

Instead of using handcrafted image features to learn a DRM,
some methods were proposed to integrate the feature learning into end-to-end nonlinear DRM learning by use of deep convolutional neural networks.
The learned end-to-end DRMs use images as their direct inputs to compute the corresponding density maps~\cite{sierra2020generating,xie2018,liu2019automated,zheng2020manifold}.
As one of the pioneering work using this strategy, Xie et al.~\cite{xie2018} proposed a fully convolutional regression network (FCRN) to learn such a DRM integrating image feature extraction and density map estimation for arbitrary-sized input images.
By use of CNNs in feature extraction and model learning, their method demonstrated superior cell counting performance than conventional density regression-based methods, especially on microscopy images containing severely overlapped cell regions.
Following Xie et al.'s work, Zheng et al.~\cite{zheng2020manifold} trained a FCRN by incorporating a manifold regularization based on the graph Laplacian of the estimated density maps
to reduce the risk of overfitting.
Liu et al.~\cite{liu2019novel} employed a post-processing CNN to further regress the estimated density map to improve the accuracy of cell counting.

However, in the original FCRN work, the network layers of a FCRN are structured hierarchically and the output of each layer relies merely on the output of its direct adjacent layer. This restricts the FCRN to produce a more authentic density map for cell counting.
In addition, the training of original FCRN is based on a single loss that is measured at the final output layer, and all its intermediate layers are optimized based on the gradients back-propagated from this single loss only. The decreased gradients potentially trap the optimization of intermediate layers into unsatisfying local minima and jeopardize the overall network performance.

Recently, CNNs that concatenate multi-scale features by shortcut connections of non-adjacent layers have been reported and demonstrated promising performance than conventional hierarchical networks for many applications~\cite{ronneberger2015,dong2017automatic}.
In these concatenated network architectures, the multi-scale image features extracted by all the layers
along the down-sampling path can be integrated into the input of the layers along the up-sampling path
to further improve the model performance.
Also, deeply-supervised (or deep supervision) learning strategies,
aiming at enhancing the training of intermediate layers of designed neural networks
by providing direct supervisions for them, have been proposed and have yielded promising performance for several computer vision tasks including image classification~\cite{lee2015deeply} and segmentation~\cite{zeng20173d,dou20173d}.
To the best of our knowledge, deeply-supervised learning has not been employed in learning a density regression model for cell counting task except our preliminary work~\cite{he2019automatic}.

In this study, a novel density regression-based method for automatically counting cells in microscopy images is proposed.
It addresses the two shortcomings that exist in the original FCRN by integrating the concatenation design and deeply-supervised learning strategy into the FCRN.
Specifically, the density regression model (DRM) is designed as a concatenated FCRN (C-FCRN) to employ multi-scale image features for the estimation of cell density maps from given images.
The C-FCRN can fuse multi-scale features and improve the granularity of the extracted features
to benefit the density map regression.
It also facilitates the learning of intermediate layers in the down-sampling path by back-propagating the gradients conveyed via the shortcut connections.
In addition, auxiliary convolutional neural networks (AuxCNNs) were employed to assist in training the C-FCRN
by providing direct and deep supervision on learning its intermediate layers to improve the cell counting performance.

The remainder of the manuscript is organized as follows.
The proposed automatic cell counting method is described in Section~\ref{sec:methodology}.
Section~\ref{sec:data} describes the testing datasets and the implementation details of the proposed method.
Section~\ref{sec:experiment} contains the experimental results.
A discussion and conclusion are provided in Section~\ref{sec:discussion} and Section~\ref{sec:conclusion}, respectively.

\section{The Proposed Cell Counting Method}
\label{sec:methodology}

\subsection{Background: Density regression-based cell counting}
\label{ssec:background}

The salient mathematical aspects of the density regression-based counting process can be described as below. For a given two-dimensional microscopy image $X\in \mathbb{R}^{M\times N}$ that includes $N_c$ cells, the density map corresponding to $X$ can be represented as $Y\in\mathbb{R}^{M\times N}$. Each value in $Y$ represents the number of cells at the corresponding pixel of $X$.
Let $\phi(X)$ be a feature map extracted from $X$, a density regression function $F_{\phi}(\phi (X), \Theta)$ can be defined as a mapping function from $X$ to $Y$:
\begin{equation}
	Y = F_{\phi}(\phi(X);\Theta),
	\label{eq:estimatemap}
\end{equation}
where the vector $\Theta$ parameterizes $F_{\phi}$.
The number of cells in $X$ can be subsequently computed by:
\begin{equation}
	N_c = \sum_{i=1}^{M}\sum_{j=1}^{N} Y_{i,j} = \sum_{i=1}^{M}\sum_{j=1}^{N} [F_{\phi}(\phi(X);\Theta)]_{i,j},
	\label{eq:counting}
\end{equation}
where $[F_{\phi}(\phi(X);\Theta)]_{i,j}$ is the computed density associated with the pixel $X_{i,j}$. The key component of density regression-based methods is to learn $F_{\phi}(\phi (X), \Theta)$ from $\phi(X)$ and the corresponding $\Theta$~\cite{lempitsky2010,fiaschi2012}.
In the fully convolutional regression network (FCRN)~\cite{xie2018},
$F_{\phi}(\phi (X), \Theta)$ can be simplified to $F(X, \Theta)$ because it can be learned directly from $X$.

\subsection{Concatenated FCRN-based cell counting method}
\label{sssec:drm}

\begin{figure}[h]
	\centering
	\includegraphics[width=0.8\textwidth]{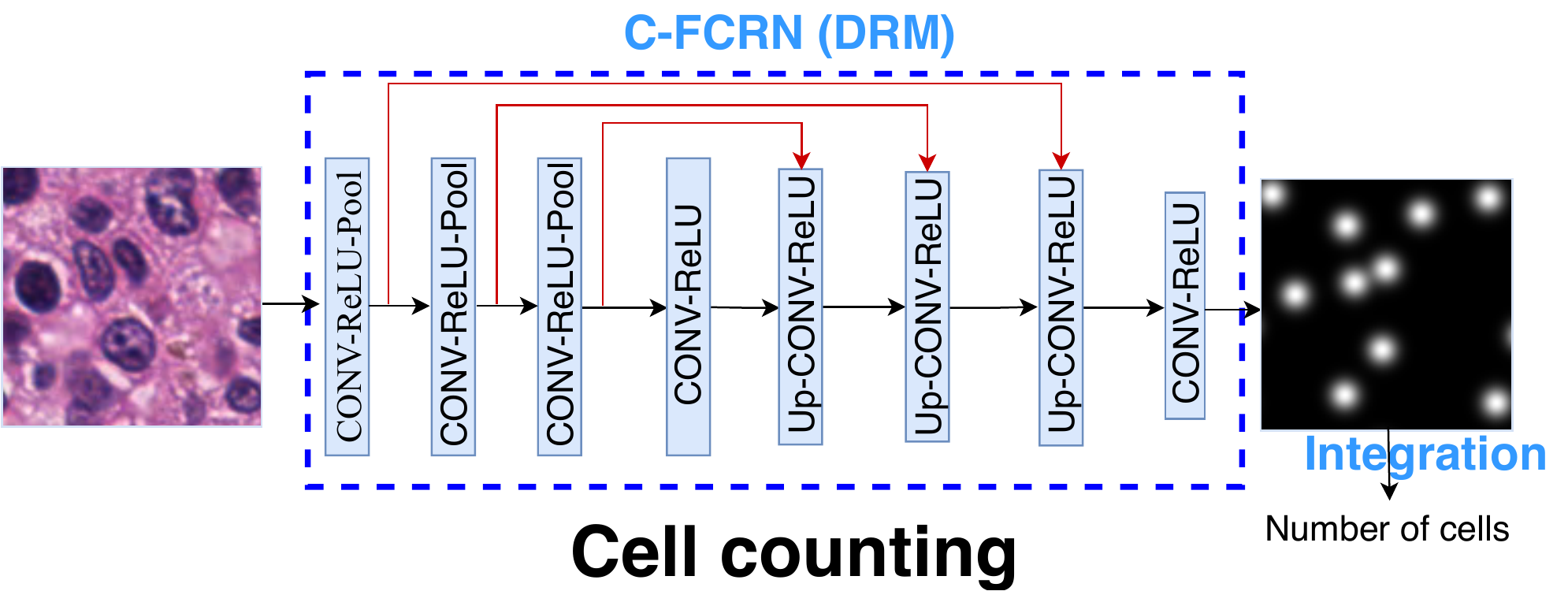}
	\caption{Framework of the proposed C-FCRN based automatic cell counting method. Different from the original FCRN, three shortcut connections (red lines) are established to connect the first three intermediate blocks to the fifth to seventh blocks, respectively.}
	\label{fig:framework-drm}
\end{figure}

The proposed concatenated FCRN (C-FCRN) is shown in Figure~\ref{fig:framework-drm},
which integrates a concatenated neural network design and deeply-supervised learning strategy into the original FCRN.
The C-FCRN network includes 8 blocks.
Three concatenation layers (red lines in Figure~\ref{fig:framework-drm}) are established
to connect the intermediate outputs along the down-sampling path
to the input of the fifth to seventh blocks along the up-sampling path, respectively.
This C-FCRN design integrates multi-scale features from non-adjacent layers to improve the granularity of the extracted features for density map regression, and subsequently improve the model performance on cell counting.
The first three blocks in the C-FCRN are employed to extract low-dimension feature maps.
Each of them includes a convolutional (CONV) layer, a ReLU layer, and a max-pooling (Pool) layer.
The fourth block, including a CONV layer and a ReLU layer, is used to further extract highly-representative features.
The fifth to seventh blocks are employed to gradually restore the resolutions of feature maps while refining the extracted feature maps.
Each of these blocks includes an up-sampling (UP) layer, a CONV layer, and a ReLU layer.
The last block, including a chain of a CONV layer and a ReLU layer, is employed to estimate the final density map.

In C-FCRN, the CONV layer in each block is associated with a set of learnable kernels and is employed to extract local features from the output of its previous layer.
The ReLU layer in each block is employed to increase the nonlinear properties of the network without affecting the receptive fields of the CONV layer by setting negative responses from its previous layer to zero while keeping the positive ones unchanged.
Each Pool layer in the first three blocks performs a down-sampling operation on an input feature map by outputting only the maximum value in every down-sampled region in the feature map.
Therefore, multi-scale informative features are extracted progressively along with the decrease of the spatial size of an input feature map. In contrast, each Up layer in the fifth to seventh block performs an up-sampling operation to gradually restore the resolution of the final estimated density map.
This network design permits integration of feature extraction into the density regression process.
Therefore, no additional feature extraction methods are required.

Given a to-be-tested image $X\in \mathbb{R}^{M\times N}$ and the trained density regression function $F(X;\Theta)$,
the density map corresponding to $X$ can be estimated as $\hat{Y} = F(X;\Theta)$.
Therefore, the number of cells in $X$ can be conveniently estimated based on the equation below:
\begin{equation}
\hat{N}_{c} =\sum_{i=1}^{M} \sum_{j=1}^{N} \hat{Y}_{i,j}=\sum_{i=1}^{M} \sum_{j=1}^{N} [F(X;\Theta)]_{i,j},
\label{eq:test}
\end{equation}
\noindent where $[F(X;\Theta)]_{i,j}$ represents the estimated density of pixel $(i,j)$ in the $X$.

\subsection{Deeply-supervised C-FCRN training with auxiliary CNNs}
\label{sssec:drm-training}

\begin{figure*}[h]
	\centering
	\includegraphics[width=0.95\textwidth]{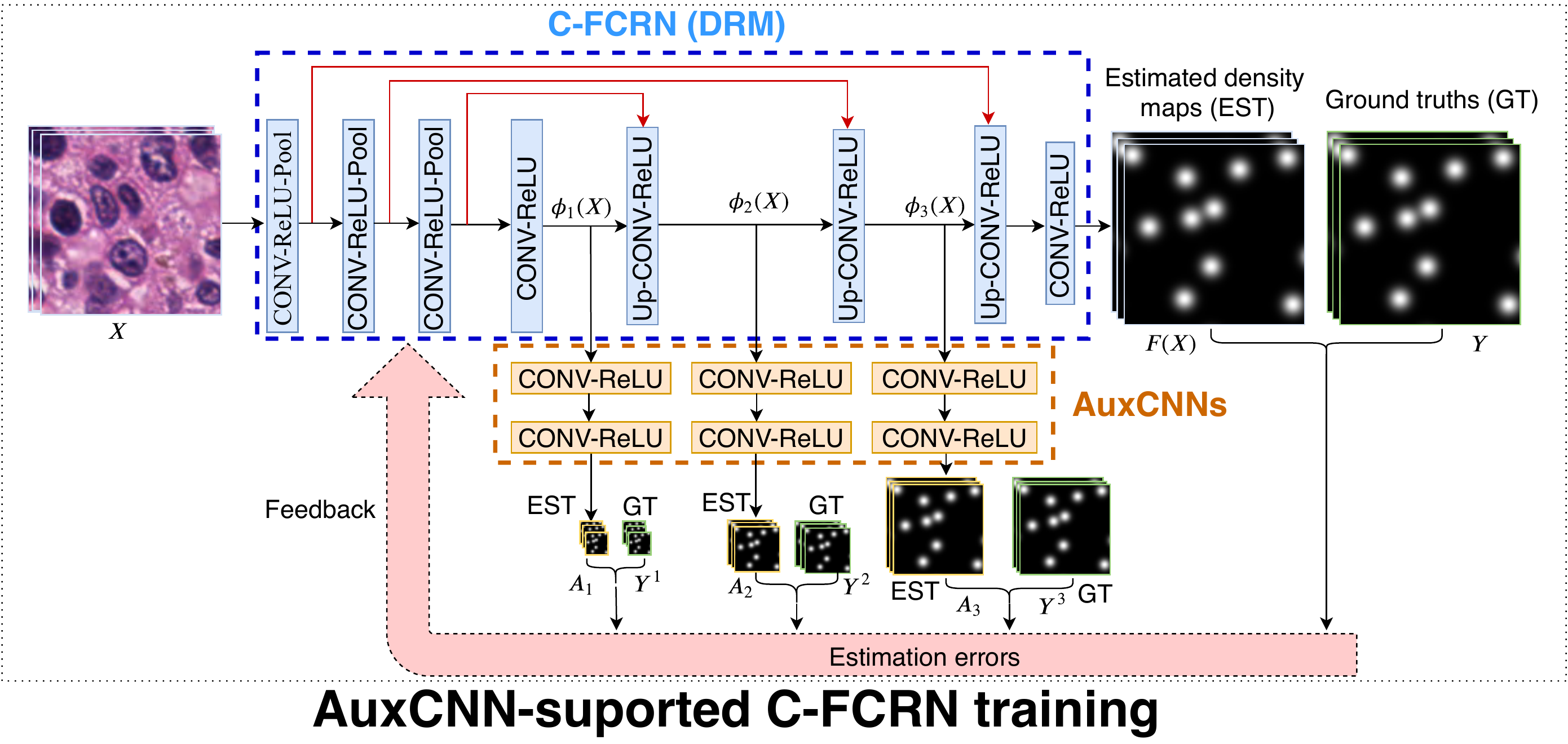}
	\caption{Framework of the AuxCNN-supported C-FCRN training process. The blue dash-line region indicates the C-FCRN.
	The orange dash-line region indicates the three AuxCNNs.
	EST and GT represents the estimated and ground truth density maps with varied resolutions, respectively.
	}
	\label{fig:framework}
\end{figure*}

The task of training the C-FCRN corresponds to learning a nonlinear density regression function $F(X, \Theta)$ with parameters $\Theta$.
However, training such a hierarchical and concatenated deep neural network by solving the corresponding highly non-convex optimization problem is a challenging task.
Motivated by the deeply-supervised learning strategies~\cite{lee2015deeply,zeng20173d,dou20173d}, we employed three auxiliary convolutional neural networks (AuxCNNs) to provide direct supervision for learning the intermediate layers of the C-FCRN.
The AuxCNN-supported C-FCRN training process is shown in Figure~\ref{fig:framework}.
Each AuxCNN contains two CONV-ReLU blocks, which estimate a low-resolution density map from each input feature map, respectively.
The difference between the estimated density maps and the related ground truth are employed
to support the C-FCRN training.

The $\Theta$ in the density regression function $F(X, \Theta)$ can be re-defined as $\Theta = (\Theta_1, \Theta_2, \Theta_3, \Theta_4)$, in which
$\Theta_1$ represents the trainable parameters in the first four blocks,
$\Theta_2$ represents the parameters in the $5$-th block,
$\Theta_3$ represents the parameters in the $6$-th block,
and $\Theta_4$ represents the parameters in the last $7$-th and $8$-th blocks, respectively.
The outputs of the $4$-th, $5$-th, and $6$-th blocks can then be denoted as
$\phi_1(X;\Theta_1)$, $\phi_2(X;\Theta_1,\Theta_2)$, and $\phi_3(X;\Theta_1,\Theta_2,\Theta_3)$.
They are also the inputs of the $1$-st, $2$-nd, and $3$-rd AuxCNNs, respectively.
Given each input $\phi_k\,(k = 1,2,3)$, the output of each AuxCNN is a low-resolution density map $A_k(\phi_k;\theta_k)$,
where $\theta_k$ represents the parameter vector of the $k$-th AuxCNN.

\begin{figure}[h]
	\centering
	\includegraphics[width=0.6\textwidth]{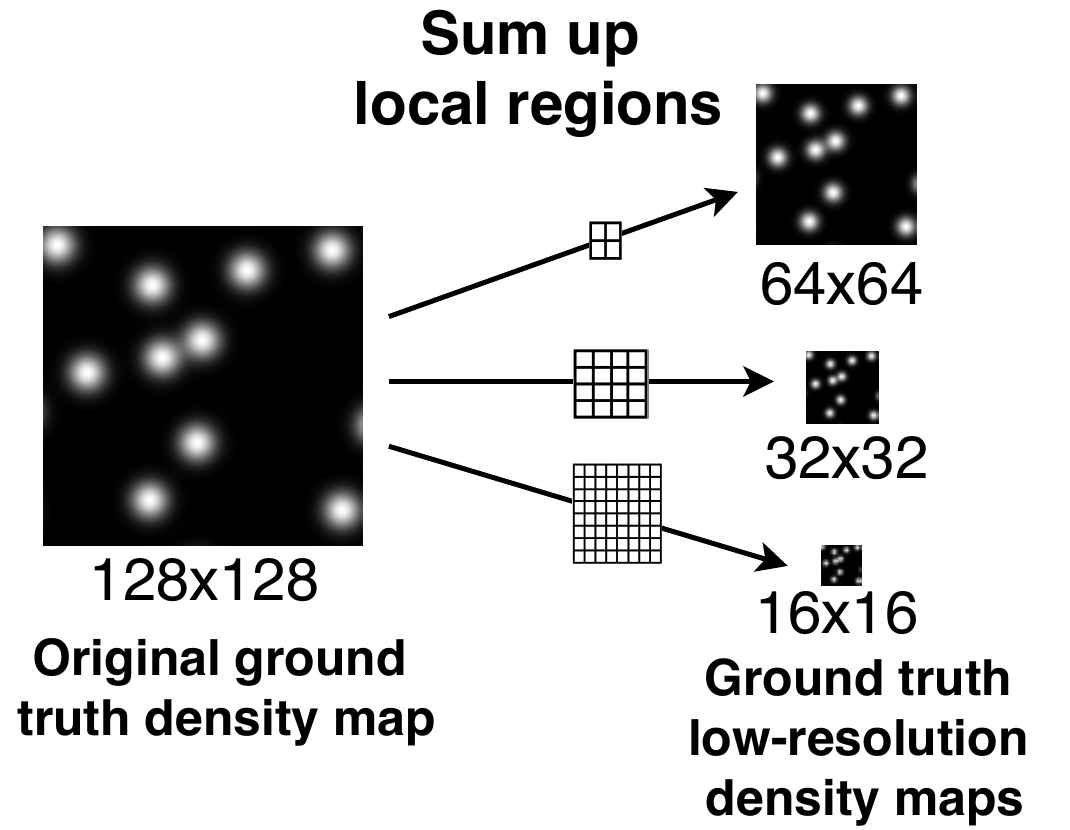}
	\caption{Example of constructing ground truth low-resolution density maps from an original ground truth of $128\times128$ pixels by summing up every local regions with size $2\times2$, $4\times4$ and $8\times 8$ pixels, respectively.}
	\label{fig:low_res_densities}
\end{figure}

$F(X;\Theta)$ and $A_k(\phi_k;\theta_k)$ are jointly trained through the minimization of a combined loss function~\cite{lee2015deeply},
\begin{equation}
\begin{aligned}
L_{cmb}(\Theta, \theta_1, \theta_2, \theta_3)
= L(\Theta)+\sum_{k = 1}^{3} \alpha_k L_k (\Theta_1, ..., \Theta_k, \theta_k) \\+\lambda (\left\| \Theta\right\|^2+\sum_{k=1}^{3}\left\| \theta_k\right\|^2), \quad k = 1,2,3,
\end{aligned}
\label{eq:loss}
\end{equation}
\noindent where $L(\Theta)$ represents a loss function
that measures the average mean square errors (MSE) between the estimated density map from the C-FCRN and the corresponding ground truth density map.
$L_k (\Theta_1, ..., \Theta_k, \theta_k)$ represents a loss function that measures the average MSE between a low-resolution density map estimated by the $k$-th AuxCNN and the corresponding low-resolution ground-truth (LRGT) density map.
The parameter $\alpha_k \in [0,1]$ controls the supervision strength under the $k$-th AuxCNN. The parameter $\lambda$ controls the strength of $l_2$ penalty to reduce overfitting and
$L_k (\Theta_1, ..., \Theta_k, \theta_k) (k = 1,2,3)$ and $L(\Theta)$ are defined as:
\begin{equation}
\label{eq:subloss}
\begin{aligned}
\begin{cases}
L_k (\Theta_1, ..., \Theta_k, \theta_k) = \frac{1}{B}\sum_{b=1}^{B} \left\| A_k (\phi_k(X_b;\Theta_1,...,\Theta_k);\theta_k) - Y^k_b \right\|^2,\\
L(\Theta) = \frac{1}{B}\sum_{b=1}^{B} \left\| F(X_b,\Theta) -Y_b \right\|^2, \quad b=1,...B,
\end{cases}
\end{aligned}
\end{equation}

\noindent where $Y_b$ represents the full-size ground truth density map of the $b$-th training data $X_b$ of $B$ training images. Here, $Y^k_b$ represents the low-resolution ground-truth (LRGT) density map,
which is generated from $Y_b$ by summing local regions in the original ground truth density map.
An example of the summing process is shown in Figure~\ref{fig:low_res_densities}.

The loss $L_{cmb}$ can be numerically minimized via momentum stochastic gradient descent (SGD) methods~\cite{bottou2010large} based on the Eqn.~(\ref{eq:update}) shown below:
\begin{equation}
\begin{aligned}
\begin{cases}
\Delta \Theta_k ^{(t+1)}  = \beta\Delta \Theta_k ^{(t)} - (1-\beta) (\eta \frac{\partial L_{cmb}^{(t)}}{\partial \Theta_k^{(t)}}),\\
\Theta_k^{(t+1)} = \Theta_k^{(t)} - \Delta \Theta_k ^{(t+1)},
\end{cases}
\end{aligned}
\label{eq:update}
\end{equation}
where $\Theta_{k}^{(t)}$ is the updated parameters $\Theta_k$ at the $t$-th iteration;
$\beta$ is a momentum parameter that controls the contribution of the result from the previous iteration;
and $\eta$ is a learning rate that determines the parameter updating speed.
Since $L_k (\Theta_1, ..., \Theta_k, \theta_k)$ only relates to $\theta_k$ and $\Theta_m\,(m = 1,2,..,k)$,
the gradient w.r.t the model parameters $\Theta_k$ can be computed by:
\begin{equation}
\frac{\partial L_{cmb}^{(t)}}{\partial \Theta_k^{(t)}} = \frac{\partial L^{(t)}}{\partial \Theta_k^{(t)}} + \sum_{m = k}^{3} \alpha_m \frac{\partial L_m^{(t)}}{\partial \Theta_k^{(t)}} +2\lambda \Theta_k^{(t)},
\label{eq:loss-mse}
\end{equation}
\noindent with the back-propagation algorithm~\cite{rumelhart1986}.
The learned $F(X;\Theta)$, represented by the trained C-FRCN model, can be used to estimate density maps for arbitrary-sized images because fully convolutional layers are employed in the C-FCRN.

In the rest of this paper, the proposed C-FCRN deeply-supervised by auxiliary CNNs during the training process is denoted as \textbf{C-FCRN+Aux}.

\section{Datasets and method implementation}
\label{sec:data}

\subsection{Datasets}
\label{ssec:testdataset}

\begin{figure}[h]
    \centering
        \includegraphics[width=0.243\textwidth]{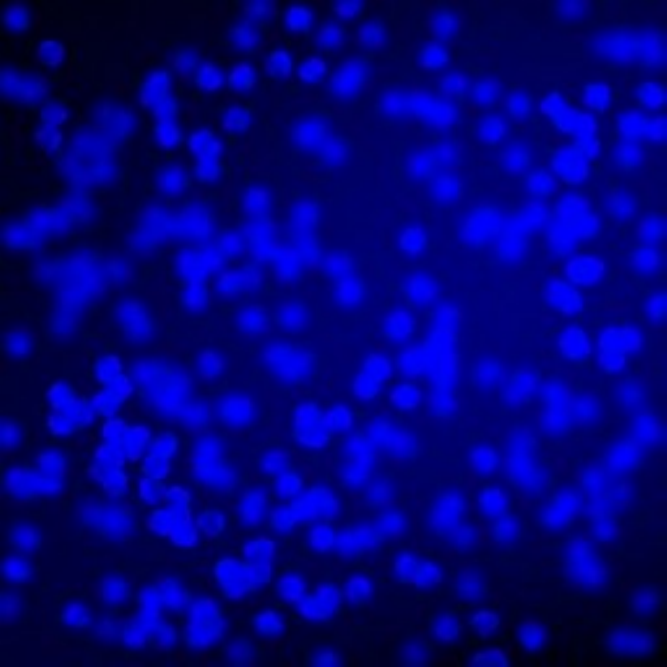}
         \includegraphics[width=0.243\textwidth]{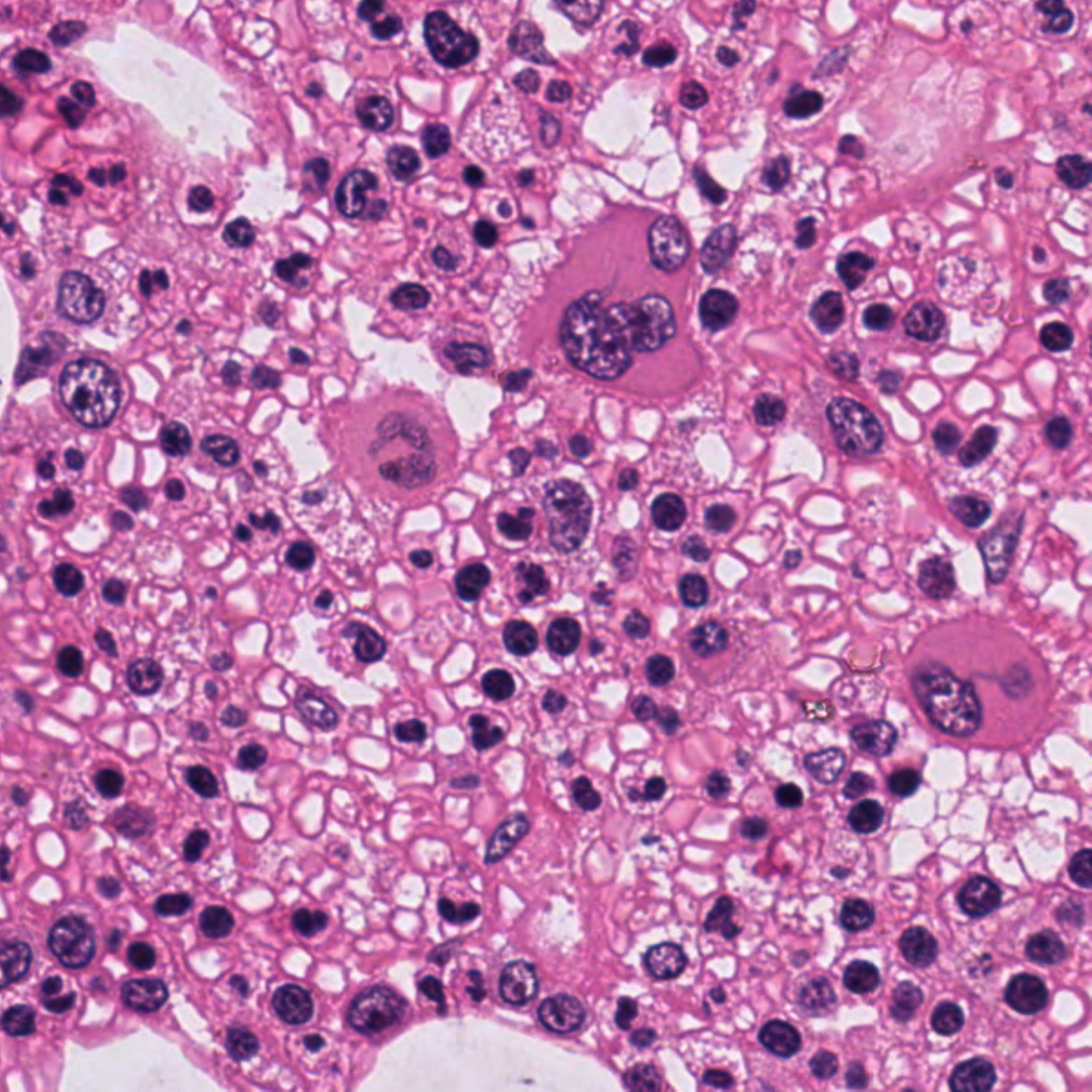}
         \includegraphics[width=0.243\textwidth]{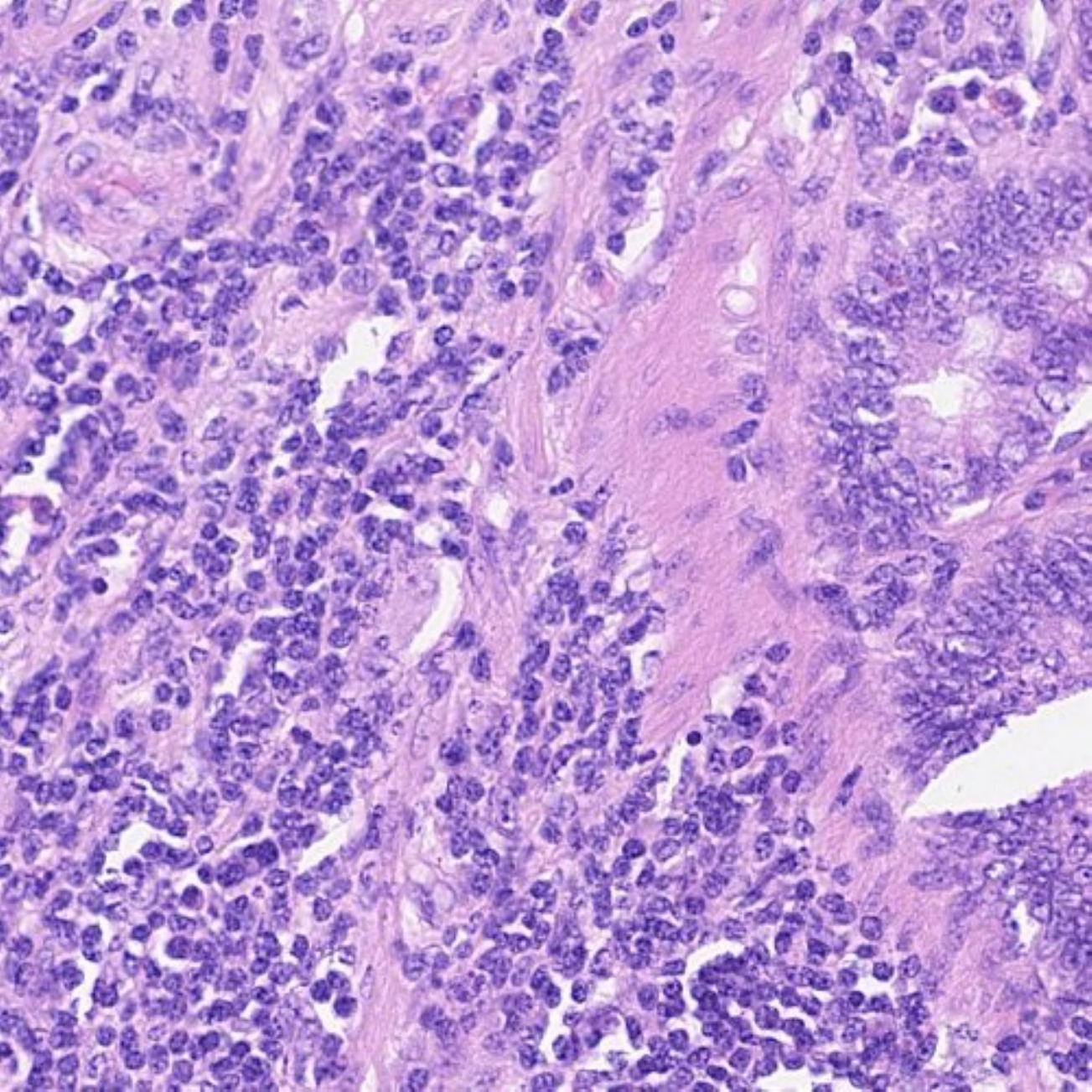}
         \includegraphics[width=0.243\textwidth]{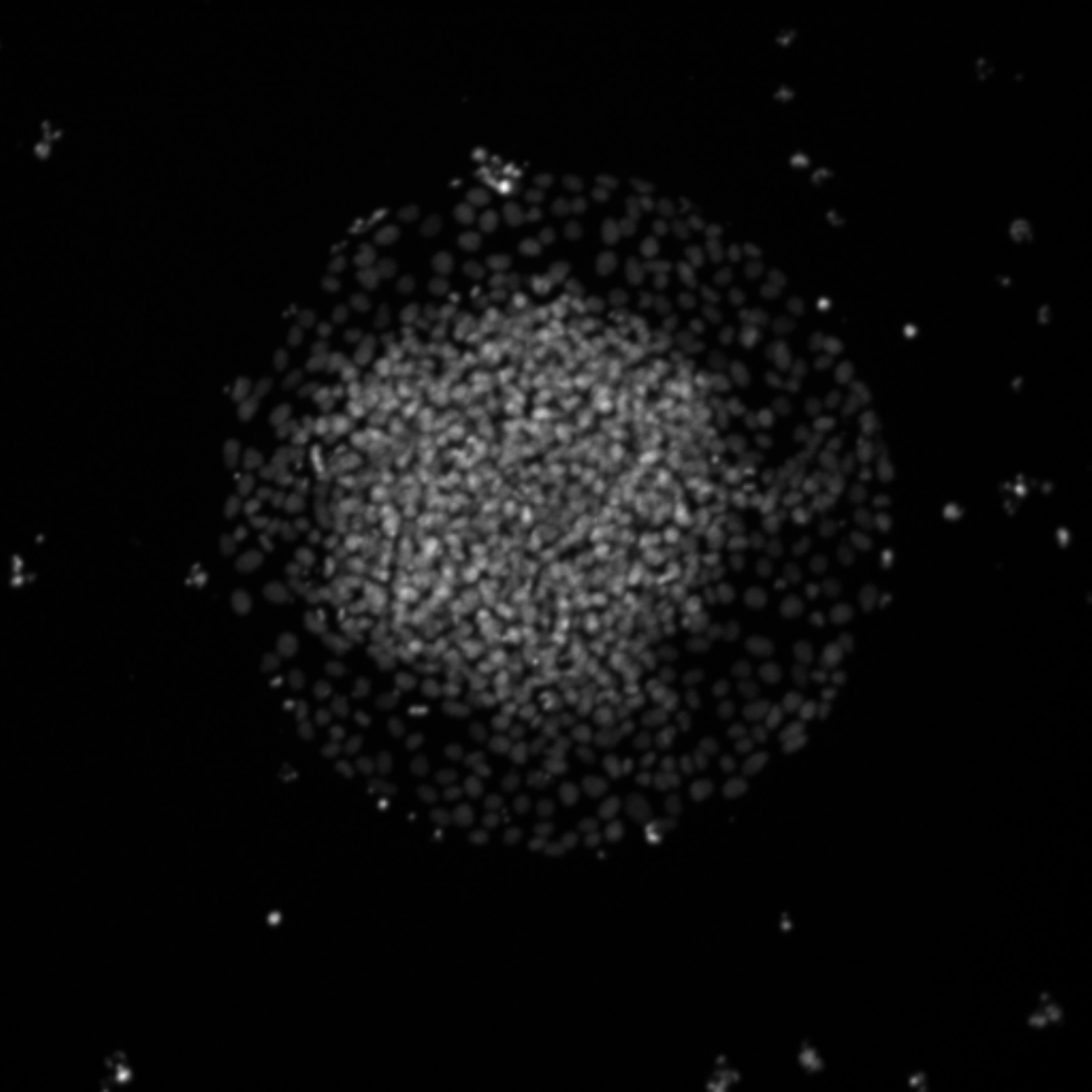}
    \caption{Example images of the four datasets used in this study.
    From left to right: Synthetic bacterial cells, Bone marrow cells, Colorectal cancer cells, and Human embryonic stem cells.}
    \label{fig:cell_samples}
    \vspace{-0.1in}
\end{figure}

Four microscopy image datasets were considered in this study, which are synthetic images of bacterial cells,
experimental images of bone barrow cells, colorectal cancer cells, and human embryonic stem cells (hESCs), respectively.
Table~\ref{tab:dataset4} illustrates the data details.
Sample images from the four datasets are shown in Figure~\ref{fig:cell_samples}.

\begin{table}[h]
	\footnotesize{
	\begin{center}
		\begin{threeparttable}
		\caption{Four datasets employed in this study}
	     \label{tab:dataset4}
            \begin{tabular}{ccccc}
            \hline\hline
            \textbf{Dataset}		& \textbf{Bacterial} & \textbf{Bone marrow}& \textbf{Colorectal}   & \textbf{hESCs} \\
           \textbf{}				& \textbf{cells } 	    	& \textbf{cells}             & \textbf{cancer cells} &  \\
            \textbf{\# of images}	& 200 						& 40 						& 100 & 49\\
            \textbf{Image size} & $256\times 256 $ 		& $600 \times 600$ 			&  $500\times 500$ & $512\times 512$\\
            \textbf{Count statistics}& $174\pm 64$		& $126\pm 33$       			&$310\pm 216$ & $518\pm316$\\
            \hline\hline
            \end{tabular}
            \begin{tablenotes}
            \item Image size is represented by pixel, and count statistics is represented by mean and standard variations of cell numbers in all the images in each dataset.
            \end{tablenotes}
            \end{threeparttable}
	\end{center}}
\vspace{-0.1in}
\end{table}

\vspace{-0.2in}

\subsubsection{Synthetic bacterial cells}
This is a public synthetic dataset generated by Lempitsky et al.~\cite{lempitsky2010}
by use of the method proposed by Lehmussola et al.~\cite{lehmussola2007computational}.
This dataset contains $200$ RGB synthetic fluorescent microscopy images of bacterial cells.
The size of each image is $256\times 256 \times 3$ pixels.
The cells in these images are designed to be clustered and occluded with each other.
This dataset is appropriate for testing the performance of the proposed method.

\subsubsection{Bone marrow cells}
This dataset includes 40 Hematoxylin-Eosin (H\&E) stained bright-field RGB microscopy images,
which were created from 10 images acquired from the human bone marrow tissues of $8$ different patients~\cite{kainz2015}.
The original image size of each H\&E image is $1200\times1200 \times 3$ pixels.
Each of the 10 original image was split into $4$ images with the size of $600\times600$ pixels, following the process in Ception-Count~\cite{cohen2017}.
The images in this dataset have inhomogeneous tissue background, and large cell shape variance.

\subsubsection{Colorectal cancer cells}
This dataset includes $100$ H\&E stained histology RGB images of colorectal adenocarcinomas acquired from $9$ patients~\cite{sirinukunwattana2016}.
Knowing the number of colorectal adenocarcinomas can help with better understanding of colorectal cancer tumor
for exploring various treatment strategies.
Images in this dataset yield high inhomogeneous tissue region, noisy background, and large variance in the numbers of cells.
This dataset is suitable to test the robustness and accuracy of given cell counting methods.

\subsubsection{Human embryonic stem cells}
This dataset contains $49$ immunofluorescent images of human embryonic stem cells (hESC) that are differentiated into varied cell types~\cite{minn2020high}.
The differentiation efficiency of the hESC population can be potentially observed based on the counted number of cells from each differentiation type in the images.
The images in this dataset yield low image contrast and severe cell occlusion and clusters.
In addition, high background noise exists in images.

\subsection{Ground truth density map generation}
\label{ssec:groundtruth}

\begin{figure}[h]
	\centering
    	\includegraphics[width=4.5in]{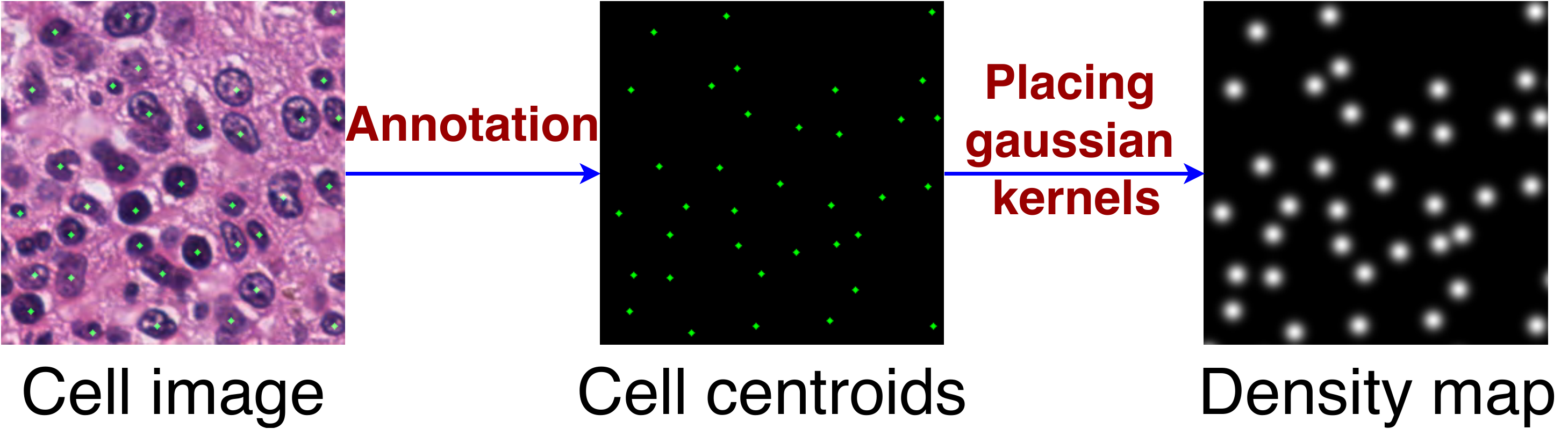}
	\caption{Example of generating density map from a given cell centroid set. }
	\label{fig:density}
\end{figure}

Both the full-size and low-resolution ground truth (LRGT) density maps of the training images need to be constructed in order to train the C-FCRN and three AuxCNNs simultaneously.
The full-size ground truth density map $Y$ of an image $X$ in the four data sets (described in Section~\ref{ssec:testdataset}) is defined as the superposition of a set of normalized 2D discrete Gaussian kernels~\cite{xie2018}.
The number of Gaussian kernels in $Y$ is identical to the number of cells in $X$,
and each kernel is centered at a cell centroid in $X$ (as shown in Figure~\ref{fig:density}). Intuitively, the density map design can be interpreted in the perspective of microscopy imaging. Due to the limitation of imaging system and the point spread function (PSF), the intensity of each single pixel in image X is affected by the PSF, and can be considered as a combination of the PSF-affected intensities of the pixel itself and its surrounding pixels. Accordingly, the density map is generated by simulating the imaging system and setting PSF as a Gaussian function.  Integrating the density over Y gives an estimate of the counts of cells in image X. This definition is also the same as the definition described in Lempitsky et al.~\cite{lempitsky2010}, one of the compared methods in this study. This process would allow density regression-based methods to solve the problem of counting the overlapping cells.
In the synthetic bacterial cell dataset, the ground truth cell centroids and numbers were automatically annotated during the image generation~\cite{lempitsky2010}, while they are manually annotated on images in the other three experimental datasets. The manual annotations for  bone marrow cell images and colorectal cell images were provided by~\cite{kainz2015} and ~\cite{sirinukunwattana2016}, respectively. The hESC annotation was performed by a graduate student under the supervision of and validation of  a biologist expert~\cite{minn2020high}.

Let $S=\{({s_{x_k}},{s_{y_k}})\in \mathbb{N}^2\}$ represent $N_c$ cell centroid positions in $X$,
where $k = 1,2, ..., N_c$.
Each $Y_{i,j}$ in $Y$ can be expressed as:
\begin{equation}
	\begin{aligned}
		\begin{cases}
		Y_{i,j} = \sum_{k=1}^{N_c} G_\sigma (i-s_{x_k},j-{s_{k_y}}), \\
		G_\sigma(n_x,n_y) = C \cdot e^{-\frac{n_x^2+n_y^2}{2\sigma^2}}, n_x, n_y \in \{-K_G, ..., 0,..., K_G\},
		\end{cases}
	\end{aligned}
	\label{eq:kernel}
\end{equation}
\noindent where $G_\sigma(n_x,n_y) \in \mathbb{R}^{(2K_G+1)\times (2K_G+1)}$
is a normalized 2D Gaussian kernel,
and $\sum_{n_x=-K_G}^{K_G}\sum_{n_y=-K_G}^{K_G} G_\sigma(n_x,n_y) =1$.
$\sigma^2$ is the isotropic covariance, \textcolor{black}{$K_G$ is an integer that determines the kernel size $(2K_G+1)\times (2K_G+1)$} pixels, and $C$ is a normalization constant.
In light of the different sizes of cells in these four different datasets, the parameter $\sigma$ was set to $5$ pixels for bone marrow images and $3$ pixels for images in the other three datasets, respectively.
The parameter $K_G$ was set to $10$ pixels for all four image datasets.

Corresponding to the bi-linear interpolation performed by the Up layers in C-FCRN,
the three low-resolution ground truth (LRGT) density maps $Y^k\in \mathbb{R}^{M_k\times N_k}(k=1,2,3)$ were generated from the original full-size ground-truth density map $Y\in \mathbb{R}^{M\times N}$ by summing local regions with size of $8\times 8$, $4\times 4$, and $2\times 2$ pixels, respectively.
Examples of ground truth of the images from the marrow bone cell dataset are shown in Figure~\ref{fig:density},
and the corresponding LRGT density map construction process is shown in Figure~\ref{fig:low_res_densities}.

All images employed in the study were preprocessed by normalizing pixel values to a uniform range $[0,1]$ in order to accelerate and stabilize the model training process~\cite{jin2015data}.
The normalized images were subsequently employed as the inputs of the networks for both training and testing purpose.
Random rotation with an arbitrary angle within $[0,40^o]$ and random flipping on the training images was performed as a data augmentation operation to mitigate overfitting.
During the training process, the ground truth density maps were amplified by $100$ in order to force the C-FCRN and AuxCNNs to fit cell area rather than the background~\cite{xie2018}.
Correspondingly, the estimated density maps estimated from the testing image were scaled back by a factor of 0.01 before counting cell numbers.

\subsection{C-FCRN and AuxCNN network parameter settings}
\label{ssec:implement}

The convolution kernel size in the first $7$ blocks of C-FCRN was set to $3\times 3$,
while that in the last block was set to $1\times 1$.
The numbers of kernels in the first to $8$-th CONV layers were set to 32, 64, 128, 512, 128, 64, 32, and 1, respectively.
The pooling size in each pool layer was set to $2\times 2$, and the Up layers performed bi-linear interpolation.
The size of the C-FCRN input image was set to $128\times 128$ pixels, so did the output density map.
Three AuxCNNs yield the similar network structures,
in which the kernel size of the first block in AuxCNN was set to $3\times3$ and the number of kernels was set to $32$,
while that in the second block were set to $1\times 1$ and $1$, respectively.

\subsection{C-FCRN+Aux training and testing}
\label{ssec:C-FCRN+Aux-training}

Six thousand epochs were employed for model training, and that can permit the convergence of the training process in this study. In each training epoch, $100$ image patches of $128\times128$ pixels were randomly cropped from each image for training.
All the cropped image patches and their corresponding density maps were employed for training DRMs in the following epoch.
The weight vector in the combined loss function $L_{cmb}(\Theta, \theta_1, \theta_2, \theta_3)$ in Eqn.~\ref{eq:loss}
was set to $(\alpha_1,\alpha_2,\alpha_3) = (\frac{1}{64},\frac{1}{16},\frac{1}{4})$, considering that the task of a higher-resolution density estimation is more correlated to the task of original density estimation task.
{A momentum SGD method was used to minimize the combined loss function for jointly training the FCRN and AuxCNNs.
The learning rates for training the C-FCRN+Aux were determined by operating a line search in a set of values $\{0.05, 0.01, 0.005, 0.0001, 0.0005, 0.001\}$ and selecting the one that results in the lowest validation loss.
Other hyper-parameters were set to the fixed values of $\beta = 0.99$, $\lambda = 0.01$, and batch size $=100$ considering the variations of these hyper-parameter values did not significantly improve the training performance based our trials.
All the network parameters in the C-FCRN+Aux were orthogonally initialized~\cite{saxe2013exact}.}

The model performance was investigated by use of 5-fold cross validation on all four image datasets.
When conducting cross validation on one of the four image datasets, the image dataset was randomly split into 5 folds of images for model training and validation.
Specifically, every time, 4 of them were employed as the training dataset and the rest one as the validation dataset.
Repeat the process for 5 times until each fold of data was used as validation dataset once.
The average validation performance over the five times were measured as the evaluation result.

The proposed C-FCRN+Aux was implemented by use of python programming language with libraries including Python 3.5, NumPy 1.14, Keras 2.0, and Tensorflow 1.3.1.
Model training and validation were performed on a Nvidia Titan X GPU with 12 GB of VRAM and several Intel(R) Xeon(R) CPUs with E5-2620 v4 @ 2.10GHz.

\subsection{Other methods for comparison}
\label{ssec:comparison}

The proposed method (denoted as C-FCRN+Aux) was compared to other eight state-of-the-art methods,
which include four regression-based counting methods~\cite{xie2018,cohen2017,lempitsky2010},
and four detection-based counting methods~\cite{arteta2016,he2017mask, xie2018efficient, falk2019u}.

Those four to-be-compared regression-based counting methods include the original FCRN method~\cite{xie2018},
the C-FCRN without AuxCNNs-supporting training (denoted as C-FCRN-only),
the Count-Ception~\cite{cohen2017} method, and the Lempitsky's method~\cite{lempitsky2010}.
The original FCRN and the C-FCRN-only methods are nonlinear density regression-based methods.
The Count-Ception~\cite{cohen2017} method is a nonlinear counter regression-based method,
which employs a fully convolutional neural network (FCNN) to perform redundant cell counting in each overlapped local region and average out the estimated results to obtain the final cell count.
The Lempitsky's method is a linear density regression-based method,
which learns the DRM by use of a regularized risk linear regression.
Its hyper-parameter settings can be found in~\cite{lempitsky2010}.

The loss functions for training the FCRN and C-FCRN were defined as the MSE between the ground truth density maps and the estimated density maps measured in a batch of training data.
Differently, the loss function in the Count-Ception method was specified as the mean absolute error between the ground truth and the estimated count maps.
The ground truth count map was generated according to its definition in the literature~\cite{cohen2017}.
A momentum SGD method was used to minimize the loss functions in all these three methods.
The learning rates and other hyper-parameters for model training in these methods were determined in the same way as those were described in Section~\ref{ssec:C-FCRN+Aux-training}.
All the network parameters in FCRN and C-FCRN-only were orthogonally initialized~\cite{saxe2013exact};
while those in the Count-Ception model were initialized by Glorot weight initialization~\cite{cohen2017}.
The local region size in the Count-Ception was set to $32 \times 32$ as suggested in the literature~\cite{cohen2017}.

The four referred detection-based counting methods include three deep-learning methods,
StructRegNet~\cite{xie2018efficient}, U-Net~\cite{falk2019u} and Mask R-CNN~\cite{he2017mask},
and the Arteta's method~\cite{arteta2016}.
In the detection-based cell counting methods, the number of cells is determined by the number of detected cell centroids or cell regions. 
The StructRegNet used a fully residual convolutional network to regress a dense proximity map that exhibits higher responses at locations near cell centroids.
Then the thresholding and non-maximum post-processes were employed to count the number of cell centroids.
Differently, the U-Net method employed a U-Net to predict a cell probability map, and count cell centroids from it.
The mask R-CNN method detects the cells by first detecting possible cell regions and then jointly predicting and segmenting these regions to get cells.
The thresholds for the post-processes were tuned by visually checking detection results for two random validation images.
The to-be-compared Arteta's method~\cite{arteta2016} aims to segment an image into non-overlapped cell regions by use of a conventional machine learning technique.
The results related to Arteta's method on the bacterial dataset was referred to the literature~\cite{arteta2016}.

The experiment settings related to the three deep learning detection-based counting methods are described as below.
The StructRegNet model was built up based on the instructions presented by Xie et al.~\cite{xie2018efficient}. The ground truth proximity map was generated by a exponential function defined as:
\begin{equation}
\label{eq:prox_map}
\mathcal{M}(u,v)=
\begin{aligned}
\begin{cases}
\frac{e^{\alpha(1-\frac{D(i,j)}{d})}-1}{e^\alpha-1}, &D(i,j)\leq d,\\
0,&D(i,j)> d,
\end{cases}
\end{aligned}
\end{equation}
where $D(i,j)$ is the Euclidean distance from a pixel $(i,j)$ to its closest annotated cell centroid; $d$ is a distance threshold and $\alpha$ is the decay ration, and both of them are used to control the shape of this exponential function. As suggested in literature~\cite{xie2018efficient}, $\alpha=3, d = 15$ was set in this study; the loss function for model training was a weighted MSE between the ground truth and estimated proximity map measured in a training batch. In this loss function, pixels closer to cell centroids were assigned to higher weights than those far-away pixels, and obtained more attention in the model training.

Although the task in this study is to annotate cell centroids,
considering that the original U-Net method~\cite{ronneberger2015} requires fully annotation of complete cell masks, we reformulated the cell counting task as a segmentation problem in order to adapt the U-Net model to infer a segmentation map containing a small 2D disk at each cell centroid for each image, as suggested by  Falk et al.~\cite{falk2019u}.
When generating the ground truth segmentation maps, the radii of the 2D disks were set to 4 pixels, 8 pixels, 5 pixels and 3 pixels for the bacterial cell, bone marrow cell, colorectal cancer cell and hESC datasets, respectively, based on the average cell size of each dataset. \textcolor{black}{The U-Net was trained by minimizing a binary cross-entropy loss with a momentum SGD method. The learning rates were determined by operating a line search in a set of values $\{0.05, 0.01, 0.005, 0.0001, 0.0005, 0.001\}$ and selecting the one that results in the lowest validation loss. 
Other hyper-parameters were set to the fixed values of $\beta = 0.99$, $\lambda = 0.01$, and batch size $=100$. All the network parameters in the U-Net were orthogonally initialized.} 
The same adaptation was performed for the Mask R-CNN method, except that a separate segmentation map was generated for each cell.
For example, a set of $N_c$ separate segmentation maps were prepared as the ground truth for an image containing $N_c$ cells. ResNet-101 was chosen as feature extraction network in the Mask R-CNN model, since it yields better performance than the ResNet-50. \textcolor{black}{The image scaling factor parameter was set to $2$. The model was trained with image patches of $512\times512\times3$ pixels that were randomly cropped from the scaled images in the training mode, and then tested on the whole scaled images.
The sizes of anchors related to the region proposal networks for the bacterial cell dataset and the bone marrow cell dataset were set to \{8, 16, 32, 64\} and \{8, 16, 32, 64, 128\}, respectively.
The Mask R-CNN model was trained by jointly minimizing the bounding box loss, classification loss, and segmentation loss. 
A stochastic gradient descent method was employed to minimize the losses. 
The batch size and learning rate were set to $4$ and $0.001$, respectively.
The other parameter settings can be found in the repository~\cite{abdulla2017mask}.}

The implementations of the six to-be-compared deep learning-based methods, including the FCRN, C-FCRN-only, Count-Ception, U-Net, Mask R-CNN, and StructRegNet, were based on the same Python, Tensorflow and Keras libraries as described in Secion~\ref{ssec:C-FCRN+Aux-training}.
In addition, the buildup of Mask R-CNN model was based on an open-sourced repository~\cite{abdulla2017mask}.
A Matlab implementation of Lempitsky's method provided by Lempitsky et al.~\cite{lempitsky2010} was used to evaluate the Lempitsky's method.
The results related to Arteta's method on the bacterial dataset was directly referred to the literature~\cite{arteta2016}.

\subsection{Performance evaluation metrics}

Mean absolute count error (MAE), mean relative count error (MRE),
and their related standard deviations (denoted by STDa and STDr) were employed as the evaluation metrics:
\begin{equation}
 \begin{array}{l}
\text{MAE} = \frac{1}{T}\sum^{T}_{t=1}{\lvert} N_{c_{t}} - \hat{N}_{c_{t}}{\rvert}, \\
\text{STDa} = \sqrt{\frac{1}{T-1}\sum^{T}_{t=1}({\lvert} N_{c_{t}} - \hat{N}_{c_{t}}{\rvert}-\text{MAE})^2},\\
\text{MRE} = \frac{1}{T} \sum^{T}_{t=1} \frac{ {\lvert} N_{c_{t}}-\hat{N}_{c_{t}} {\rvert}}{N_{c_{t}}},\\
\text{STDr} = \sqrt{\frac{1}{T-1}\sum^{T}_{t=1}(\frac{ {\lvert} N_{c_{t}}-\hat{N}_{c_{t}} {\rvert}}{N_{c_{t}}}-\text{MRE})^2}.\\
 \end{array}
\end{equation}
\noindent where $T$ is the number of validation images,
$N_{c_t}$ and $\hat{N}_{c_t}$ are the ground truth cell count and the estimated cell count in the $t$-th image respectively.
MAE measures the mean of the absolute errors between the estimated cell counts and their ground truths for all the validation images.
Considering the large variance in the numbers of cells in colorectal images and hESC images,
MRE was also considered for method evaluation because they measure the relative errors between the ground-truth counts and the estimated counts.
STDa and STDr indicate the stability of the cell counting process.
 A lower MAE or MRE indicates a better cell counting accuracy, and a lower STDa or STDr means a more stable counting performance.
\begin{table*}
\footnotesize{
	\begin{center}
		\begin{threeparttable}
			\caption{
				MAE$\pm$STD performance evaluated on the four data sets.
			}
			\label{tab:mae}
			\begin{tabular}{ccccc}
				\hline\hline\noalign{\smallskip}
				\textbf{\bf MAE $\pm$ STD}		& \textbf{Bacterial} & \textbf{Bone marrow}& \textbf{Colorectal} & \textbf{hESC}\\
				\textbf{}				                    & \textbf{cells } 	    	& \textbf{cells}             & \textbf{cancer cells} &  \\
				\noalign{\smallskip}
				\hline
				\noalign{\smallskip}
				Lempitsky's method &  $3.52\pm 2.99$	& --- 	& ---	&---\\
				Altera's method       &  $5.06^*$	& --- 	& ---	&---\\
				Mask R-CNN & $36.92\pm19.73$ & $44.4\pm 14.17$& ---& ---\\
				U-Net & $27.77\pm25.48$ & $48.00\pm 18.98$& ---& ---\\
				StructRegNet& $9.80\pm8.68$ & $12.75\pm 8.62$& $ 45.97\pm47.97$& $189.14\pm231.86$\\
				FCRN 							& $ 2.75\pm2.47 $& $8.46\pm 7.63 $&$42.58\pm33.51$& $44.90\pm35.39$\\
				C-FCRN-Only & $2.58\pm 2.28 $& $8.68 \pm 7.37$& $39.55\pm35.80$& $42.17\pm 30.97$\\
				Count-Ception  & $2.79\pm2.68$ & $7.89\pm 6.83$& $ 34.14\pm29.04$& $35.87\pm35.77$\\
				C-FCRN+Aux & ${\bf 2.37 \pm 2.27}$ & ${\bf 6.55\pm 5.26}$& ${\bf 29.34 \pm 25.4}$&${\bf 32.89\pm 26.35}$
				\\
				\hline
			\end{tabular}
			\begin{tablenotes}
				\item * indicates the result reported in the literature~\cite{arteta2016}, in which the method was tested on a set of 100 testing bacterial cell images. Differently, the results from other methods related to this dataset were evaluated on a complete set of 200 bacterial cell images in this study, since the cross validation-based evaluation allows each image to be considered as a testing image for once. In addition, the Lempitsky's method was only validated on the bacterial cell dataset because this dataset provides handcrafted image features for validation purpose. The results from the U-Net and Mask R-CNN were not reported on colorectal cancer cell and hESC datasets, due to their failure in providing reasonable detection results on the two datasets.
			\end{tablenotes}
		\end{threeparttable}
	\end{center}
	}
\end{table*}

\section{Experimental results}
\label{sec:experiment}

\subsection{Cell counting performance}
\label{ssec:performance}

\begin{figure}[h]
	\centering
	\includegraphics[width=0.85\textwidth]{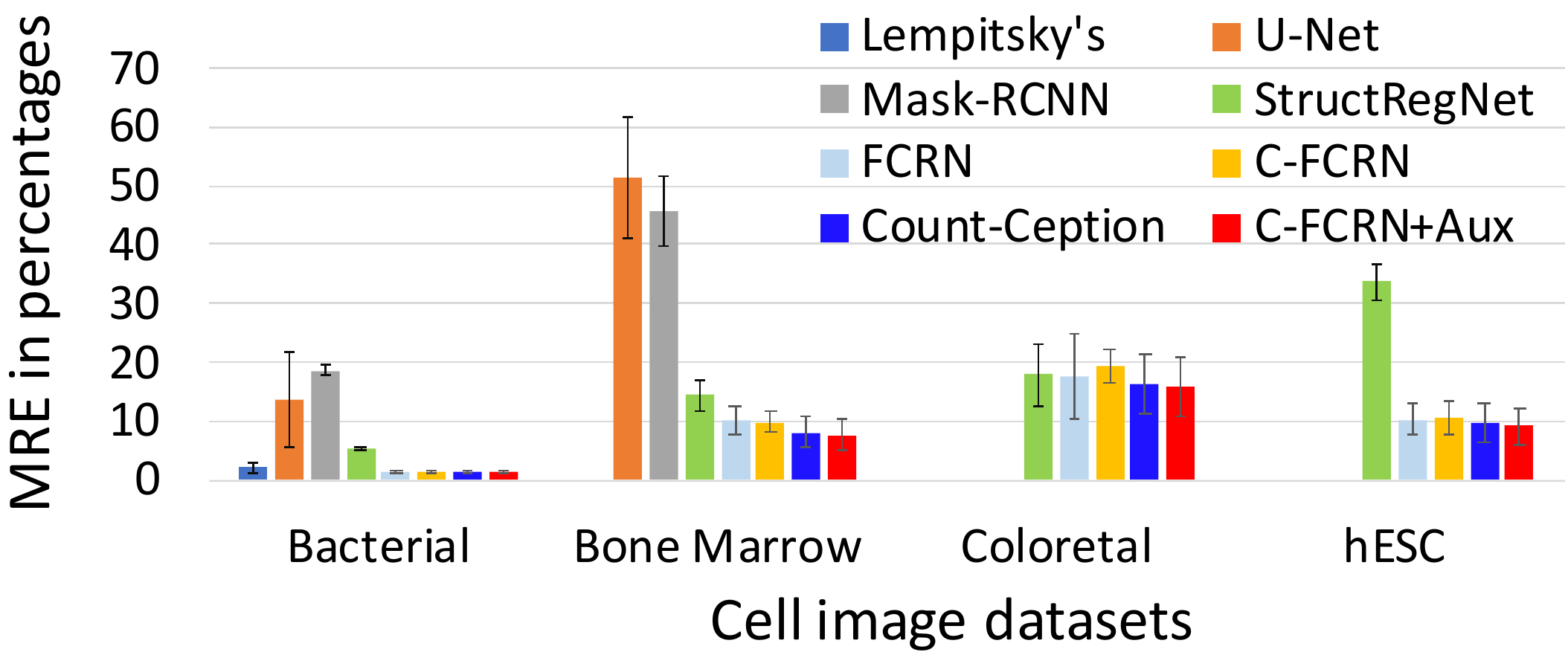}
	\caption{The MRE performance evaluated on four different datasets. ``C-FCRN+Aux'' represents the proposed method in this study.
No MRE results were reported for Arteta's method.
	}
	\label{fig:mre}
\end{figure}

Cell counting performance of the proposed ``C-FCRN+Aux'' method and the other eight methods on the four datasets are reported in the Figure~\ref{fig:mre} and Table~\ref{tab:mae}.
The proposed method demonstrates superior cell counting performance to the other eight methods in terms of MAE and MRE.
Compared to the regression-based methods, all four detection-based methods achieve worse counting performance in terms of MAE and MRE.
Also, all three non-linear density regression-based methods (the proposed method, FCRN, C-FCRN-only) demonstrate superior counting performance compared to Lempitsky's method, one of the conventional linear methods.

A paired $t$-test was performed on the absolute counting errors related to the proposed method (C-FCRN+Aux) and its closest counterpart C-FCRN-only.
In this test, the null hypothesis $H_0$ was defined as the population means of absolute errors related to the C-FCRN+Aux is higher than that of C-FCRN, and vise versa for hypothesis $H_1$.
The $p$-values for the tests on the synthetic cell, bone marrow cell, colorectal cancer cell, and hESC datasets are $6.19\times10^{-4}$, $0.042$, $5\times10^{-7}$ and $2.8\times10^{-3}$, respectively.
A similar paired $t$-test was performed on the absolute counting errors related to C-FCRN+Aux and original FCRN,
and the corresponding $p$-values related to the four datasets are $0.024$, $0.012$, $7.35\times 10^{-5}$ and $0.017$, respectively.
The paired $t$-test results show that the MAEs related to the proposed method were lower than its two counterparts: C-FCRN and FCRN-only with statistical significance.

\begin{figure*}
	\centering
	\subfigure[Bacterial cells]
	{
		\includegraphics[width=\textwidth]{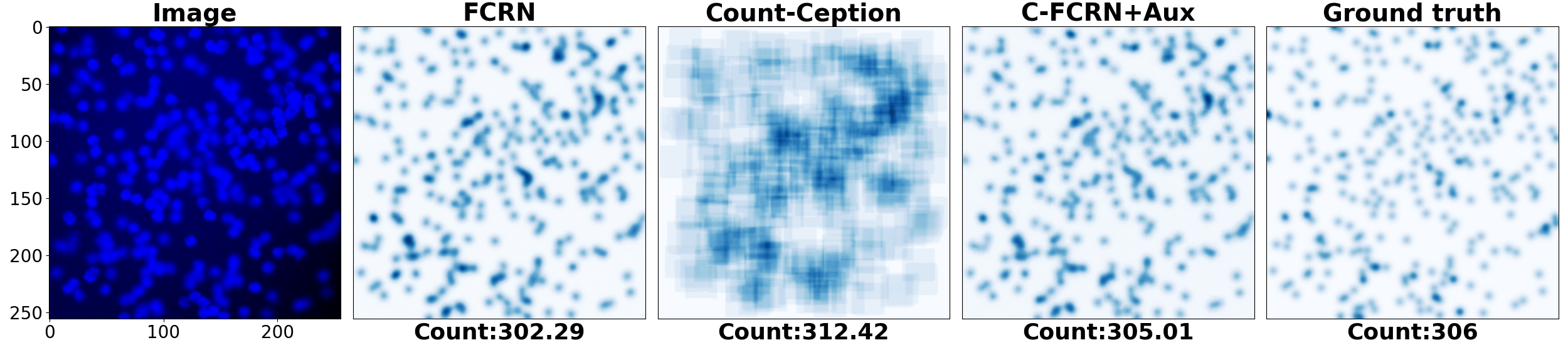}
		\label{fig:bactDen}
	}
	\subfigure[Bone marrow cells]
	{
		\includegraphics[width=\textwidth]{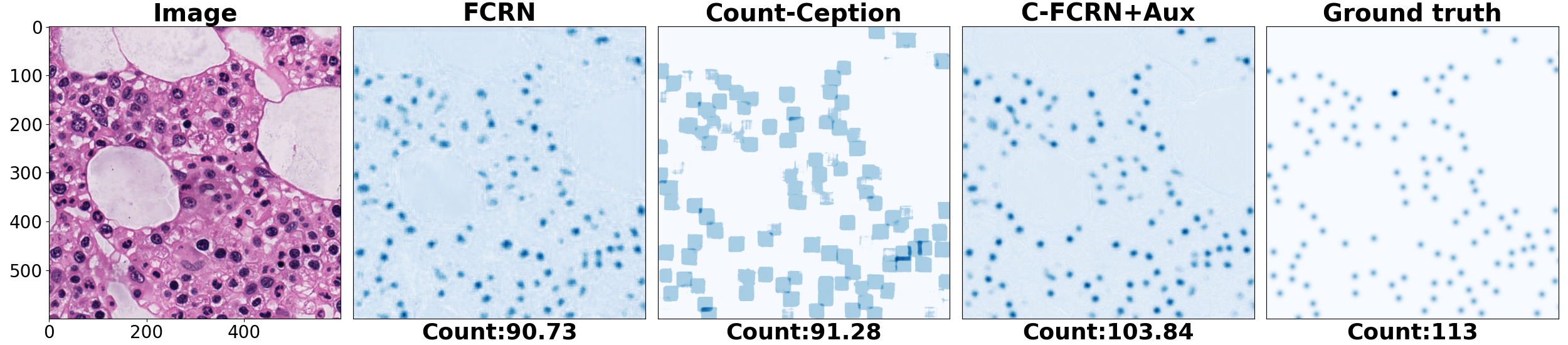}
		\label{fig:bmDen}
	}
	\subfigure[Colorectal cancer cells]
	{
		\includegraphics[width=\textwidth]{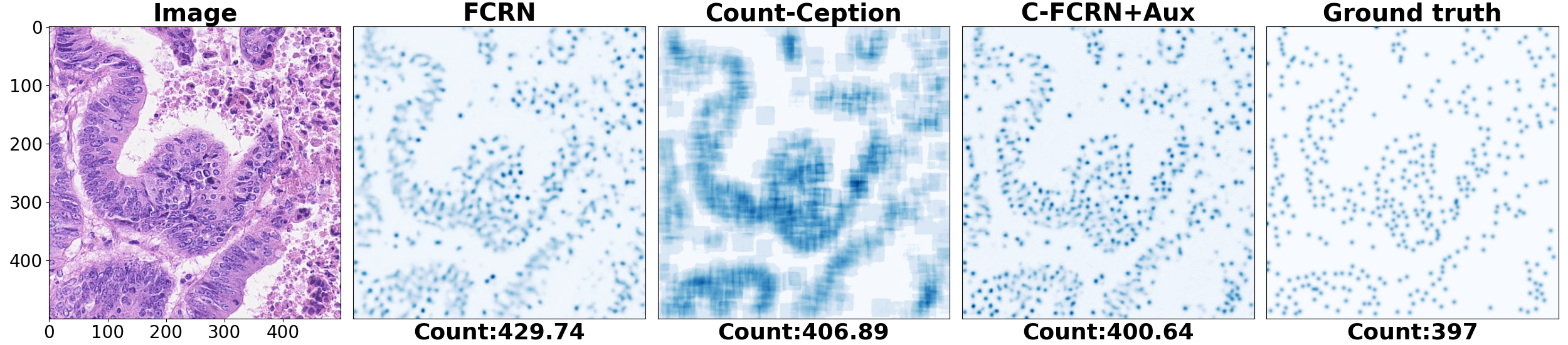}
		\label{fig:cancerDen}
	}
	\subfigure[human embryonic stem cells (hESC)]
	{
		\includegraphics[width=\textwidth]{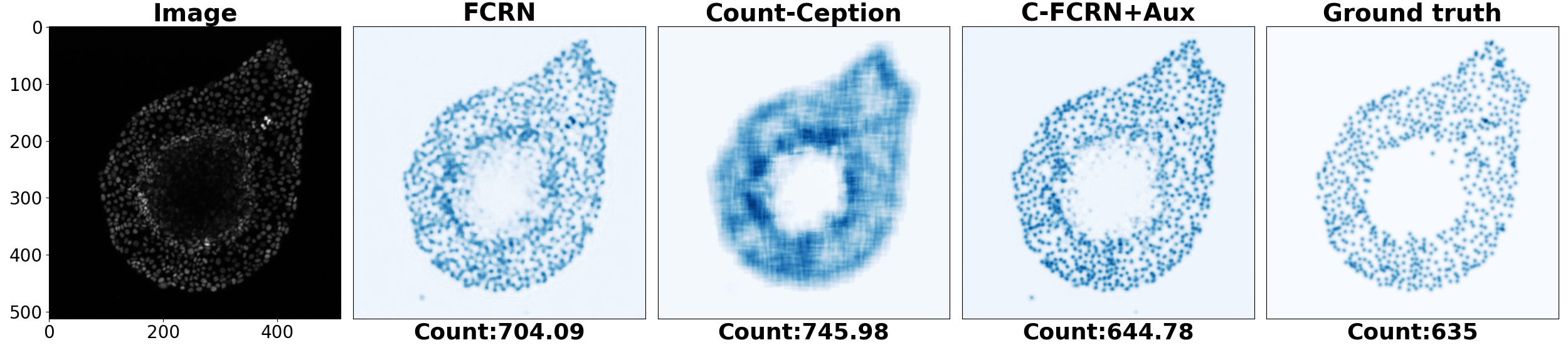}
		\label{fig:embDen}
	}
	\caption{Estimated density or count maps from a sample image in each of the four datasets.
	The panels from left to right on each row show the cell images and the density/count maps estimated by the FCRN, the Count-Ception, and the proposed method (C-FCRN+Aux), and the associated ground truth density maps, respectively.
	}
	\label{fig:densityEst}
\end{figure*}

Figure~\ref{fig:densityEst} shows the estimated density/count map of a testing example in each of the four datasets. The density maps estimated by the C-FCRN+Aux appear visually closer to the ground truth density maps compared to the FCRN method.
It is noted that the Count-Ception method predicts a count map directly without providing cell centroid locations, which is different from the other density regression-based methods.

\begin{figure}[h]
    \centering
    \subfigure[Bacterial cells (ground truth count: 306)]
    {
       \includegraphics[width=0.75\textwidth]{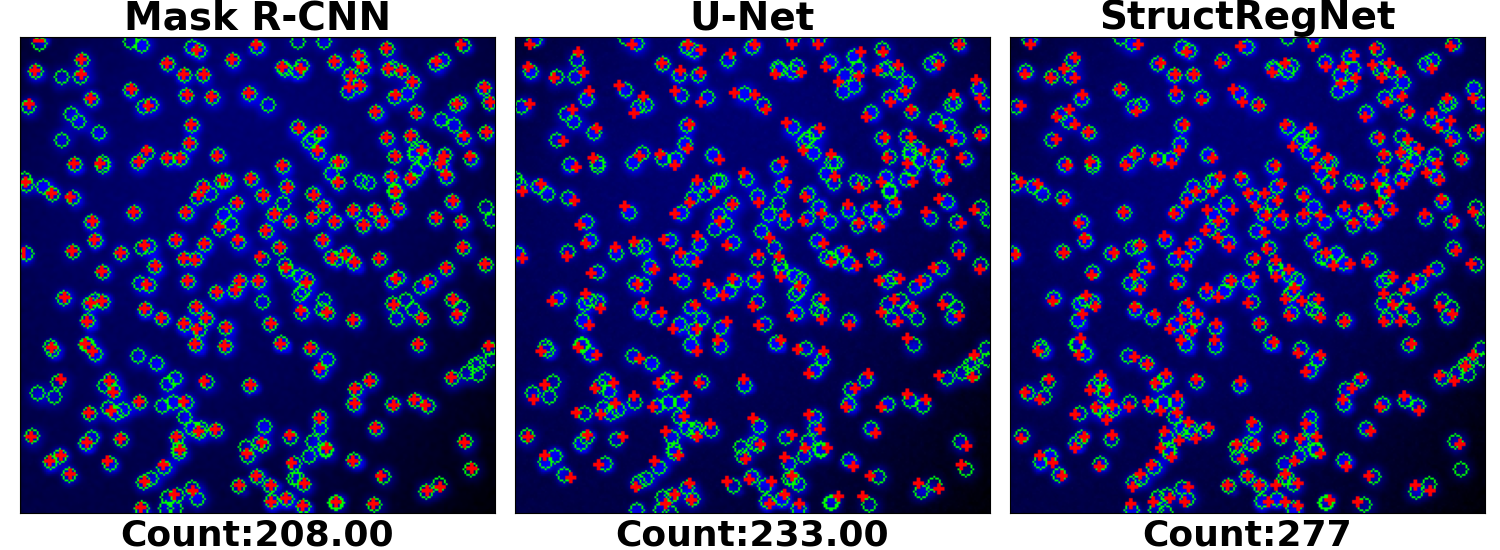}
        \label{fig:det_bacterial}
    }
    \subfigure[Bone marrow cells (ground truth count: 113)]
    {
        \includegraphics[width=0.75\textwidth]{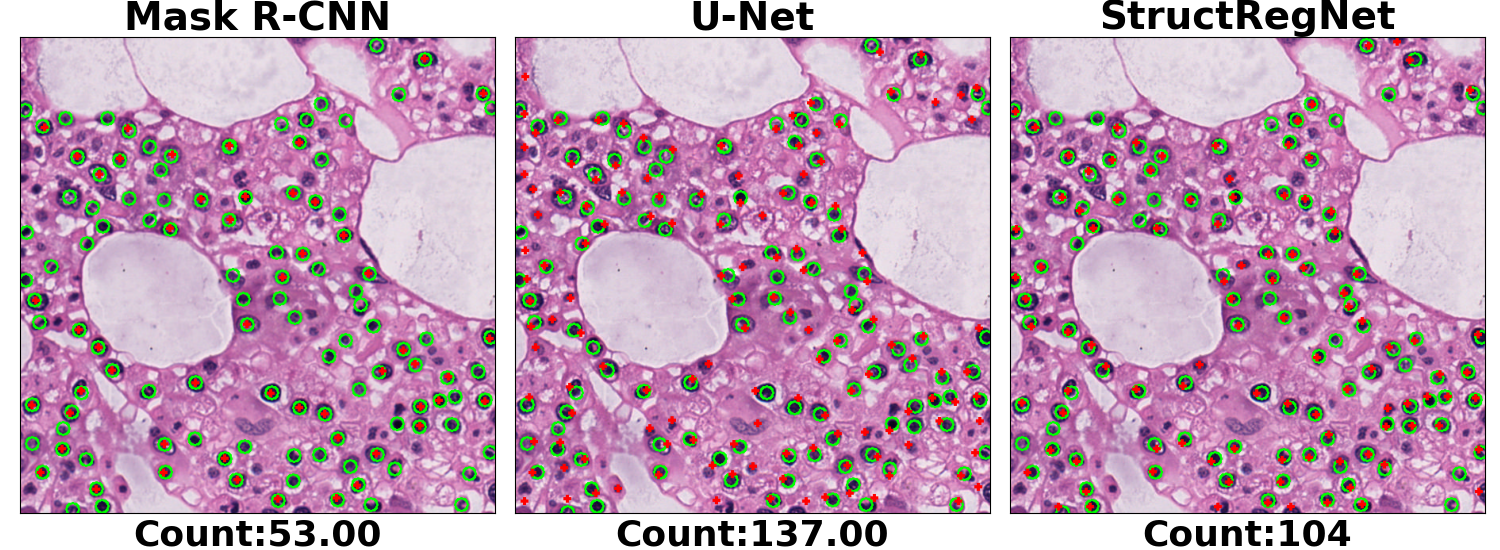}
        \label{fig:det_bm}
    }
   \caption{Example results of three deep-learning detection-based cell counting methods (Mask R-CNN, U-Net, and StructRegNet). Panels (a) and (b) show the prediction results on the bacterial and bone marrow datasets, respectively.
   The green cycles and red dots in each image represent the ground truth annotations and the detected cell centroids, respectively.}
    \label{fig:detection_results}
\end{figure}

Figure~\ref{fig:detection_results} shows the result of a testing example in each of the bacterial and bone marrow cell datasets by use of three detection-based methods (Mask R-CNN, U-Net and StructRegNet).
The StructRegNet achieves more accurate results than the other two.
One of the possible reasons is that the StructRegNet model is trained to regress a dense proximity map,
in which the pixels closer to cell centroids can get more attention than those far-away pixels;
this is different from the U-Net and Mask R-CNN model.
This can benefit more for local maximum searching in the non-maximum post-process and yield better cell detection performance.
It was also observed that the three detection-based methods commonly failed in detecting clustered and occluded cells in the bacterial image example. Also, they either under-detect or over-detect cells in the bone marrow image example. These images contain strongly heterogenous backgrounds and the shapes of cells vary largely.
The inaccuracy of cell detection with these detection-based methods confirms their lower cell counting accuracy shown in the Table~\ref{tab:mae} and Figure~\ref{fig:mre}.

\begin{figure}[h]
    \centering
    \subfigure[Colorectal cancer cells (ground truth count: 712)]
    {
       \includegraphics[width=0.75\textwidth]{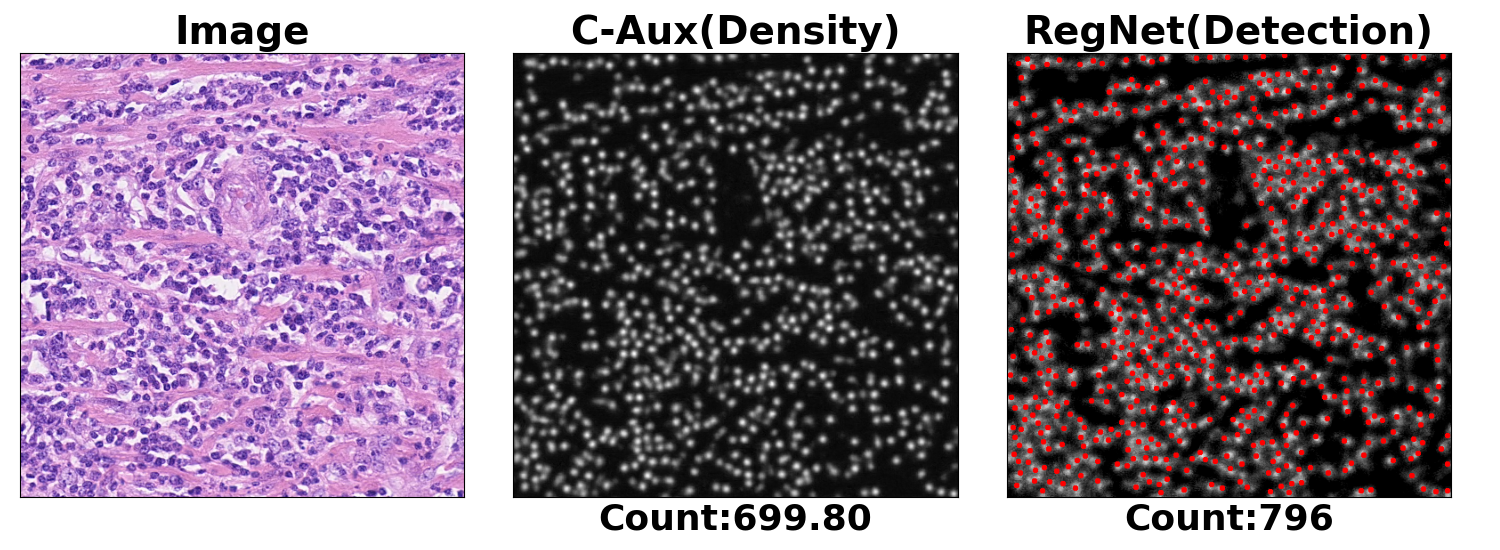}
        \label{fig:den_det_colorectal}
    }
    \subfigure[hESC (ground truth count: 1100)]
    {
        \includegraphics[width=0.75\textwidth]{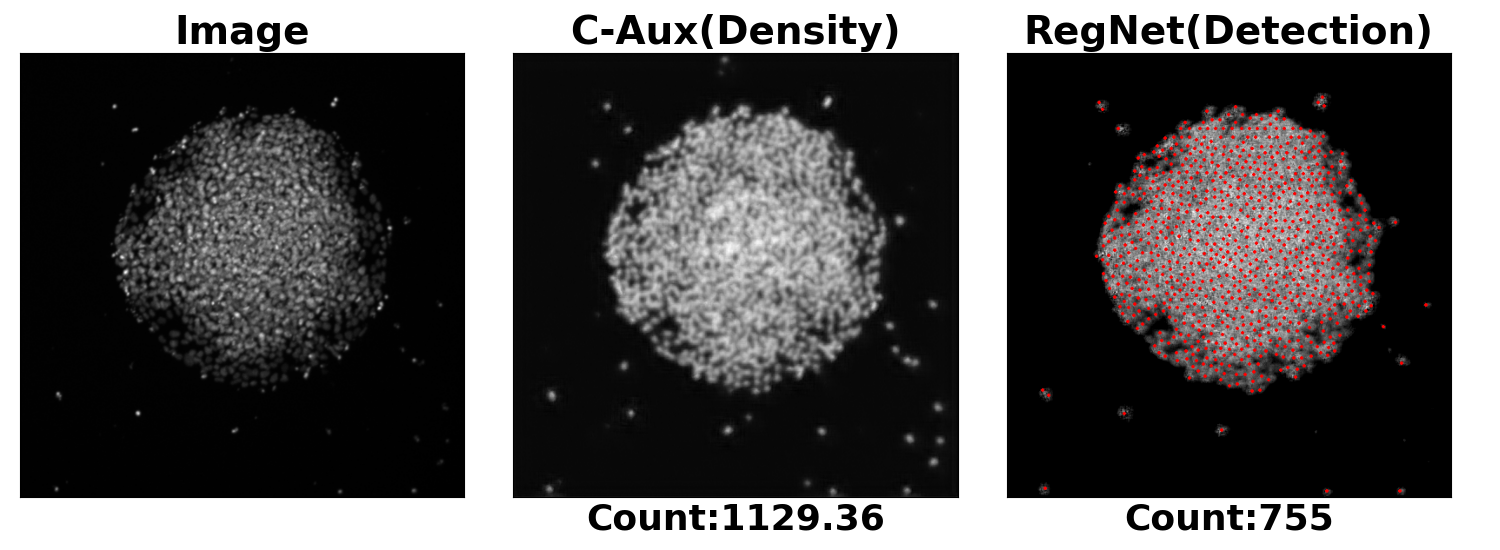}
        \label{fig:den_det_hESC}
    }
   \caption{Example prediction results based on the proposed C-FCRN+Aux method and the detection-based method (StructRegNet).  Here, ``image'', ``C-Aux'' and ``RegNet'' represent the processed image and the estimated density map using the ``C-FCRN+Aux'' method and  the computed proximity map using the ``StructRegNet'' methods. The red dots represent the detected cell centroids based on the computed proximity map, respectively.}
    \label{fig:density_det}
\end{figure}

Figure~\ref{fig:density_det} shows the result on an example in each of the colorectal and hESC datasets by use of the proposed method and the StructRegNet method, which are the best-performing regression-based method and detection-based method tested in this study, respectively.
The cells are commonly concentrated in colorectal cell images and seriously clustered and occluded in the hESC images. Cell detection in these two scenarios is extremely challenging.
The StructRegNet method shows much worse counting performance compared to the proposed method.

\subsection{Benefits of using AuxCNNs to support C-FCRN training}

\begin{figure*}[h]
	\centering
	\subfigure[Bacterial cells]
	{
		\includegraphics[width=0.45\textwidth]{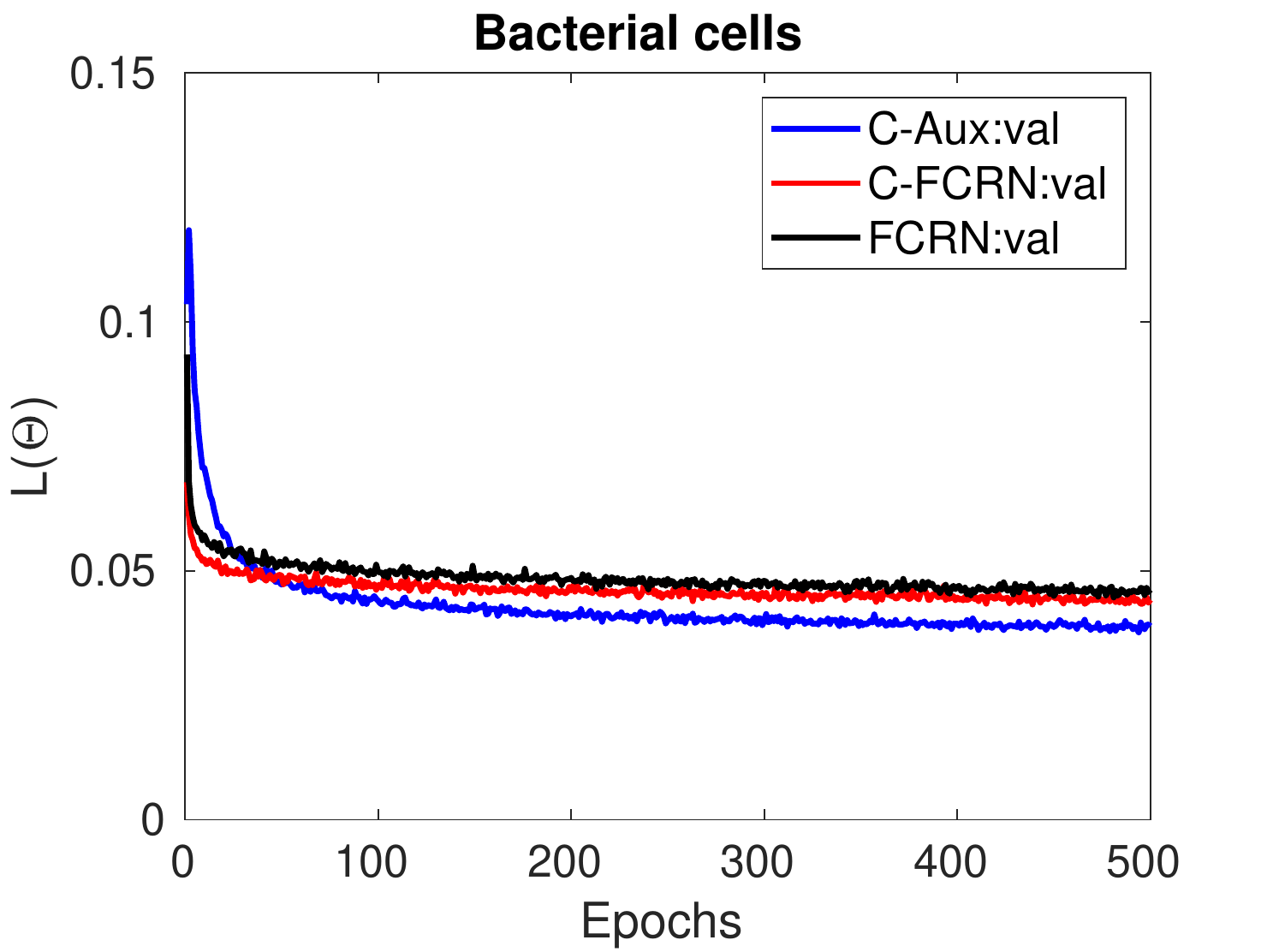}
		\label{fig:bact_conv}
	}
	\hspace{-0.2in}
	\subfigure[Bone marrow cells]
	{
		\includegraphics[width=0.45\textwidth]{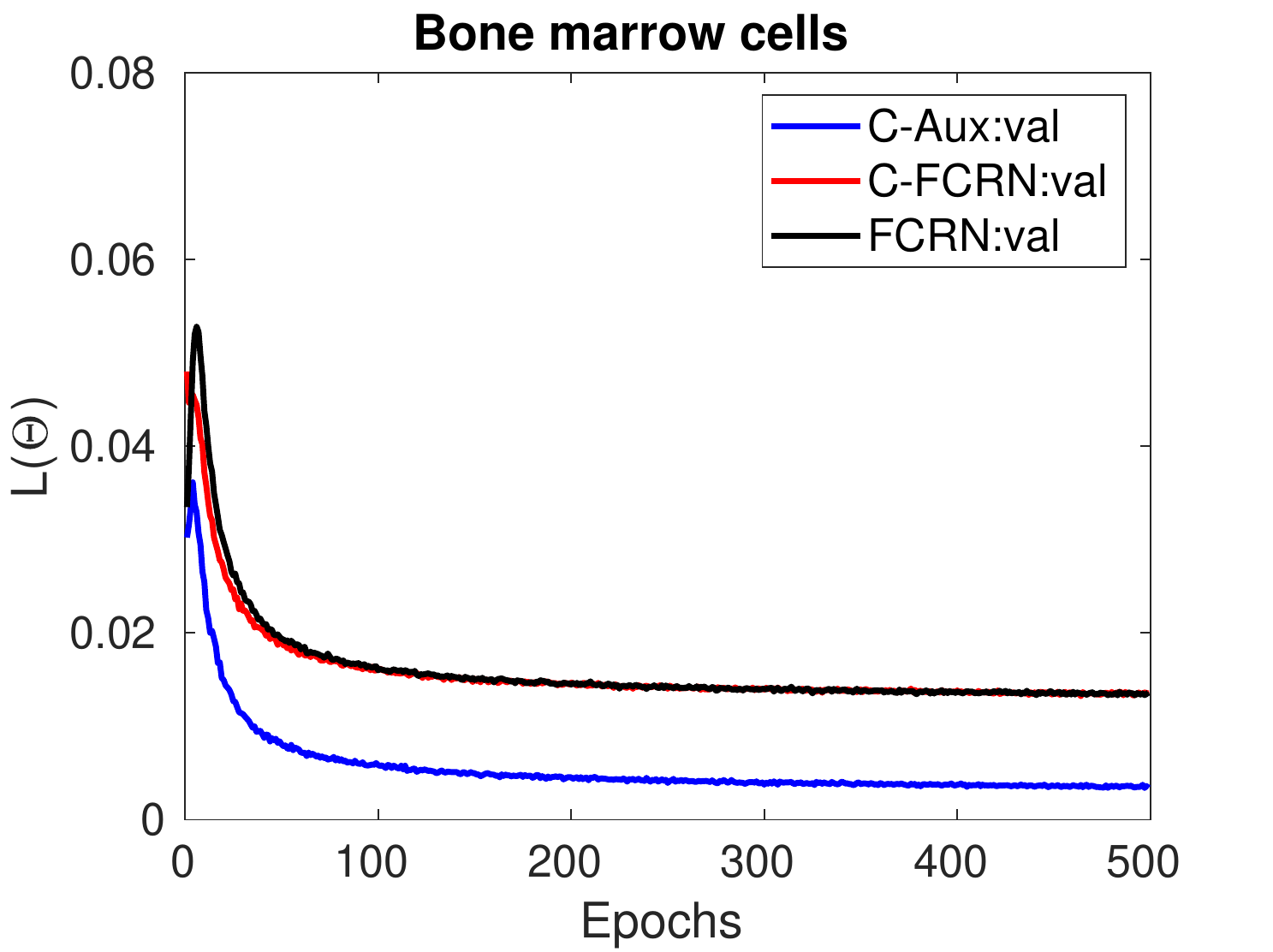}
		\label{fig:bm_conv}
	}
	\hspace{-0.2in}
	\subfigure[Colorectal cancer cells]
	{
		\includegraphics[width=0.45\textwidth]{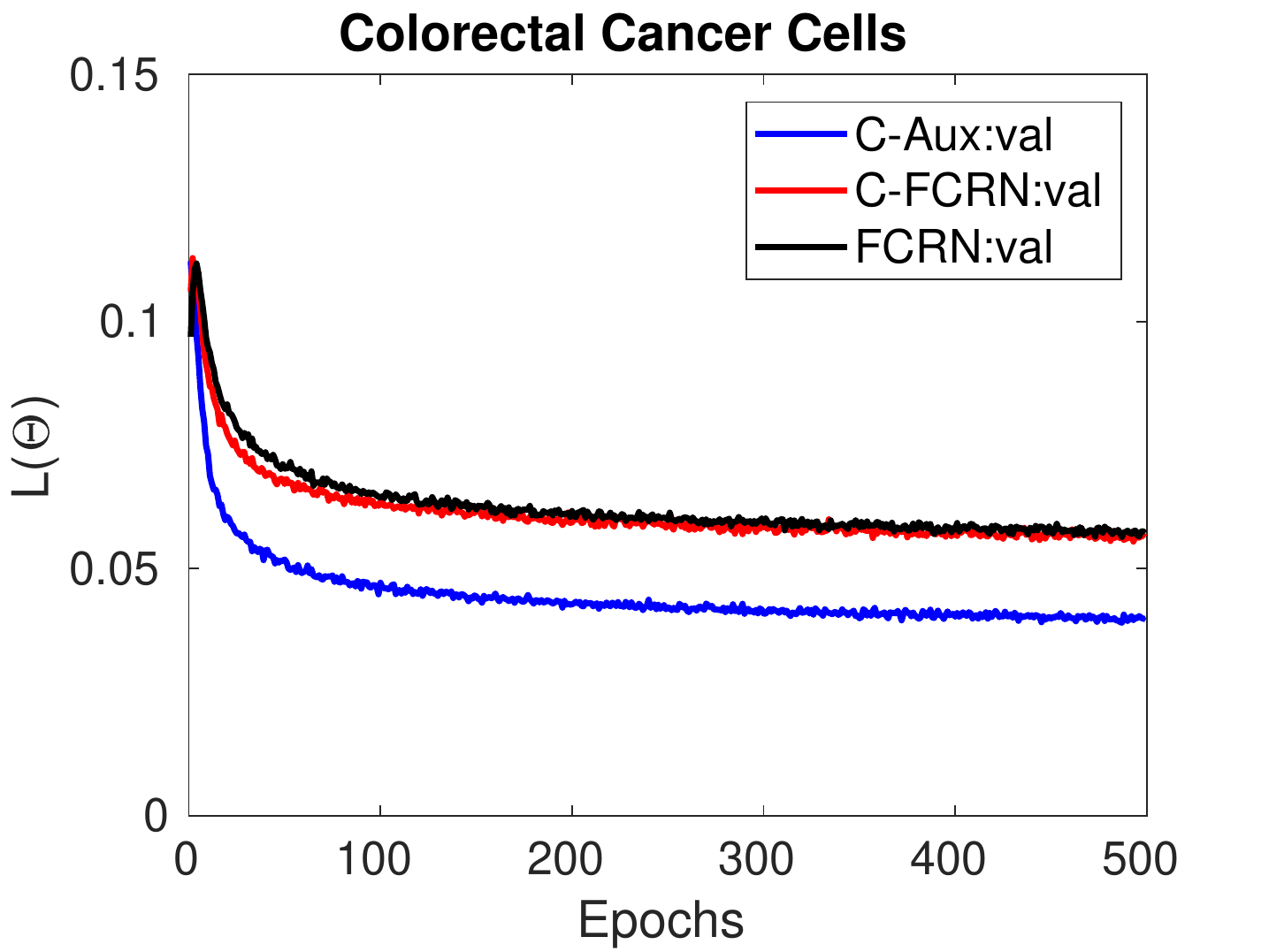}
		\label{fig:colo_conv}
	}
	\hspace{-0.2in}
	\subfigure[hESC]
	{
		\includegraphics[width=0.45\textwidth]{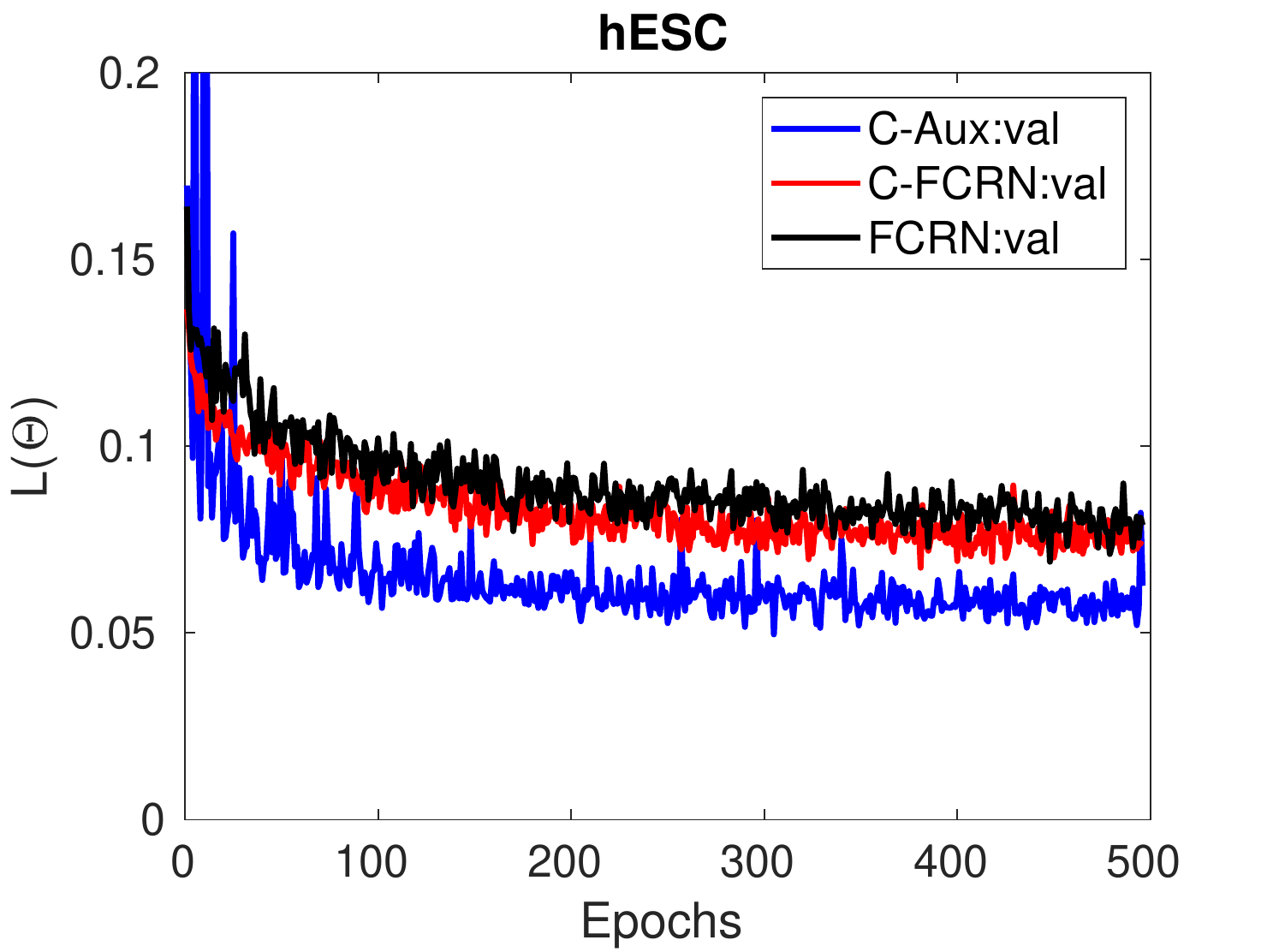}
		\label{fig:emb_conv}
	}
	\caption{Validation losses as the functions of epochs are plotted for the DRM training on the four datasets. C-Aux and C-FCRN are abbreviations of C-FCRN+Aux and C-FCRN-Only methods, respectively.}
	\label{fig:convergence}
\end{figure*}

The accuracy of the estimated density map along the training process was investigated
to demonstrate that AuxCNNs supports the overall model training process.
Figure~\ref{fig:convergence} shows the curves of validation losses vs. the number of epochs for the proposed method and the other two nonlinear density regression methods (C-FCRN-only and FCRN) on four datasets.
One of the five validation curves generated during the 5-fold cross validation procedure is presented for each method as an example.
The curves generated for the first $500$ epochs are shown because the validation losses keep stable after the $500$-th epoch.
As shown in Figure~\ref{fig:convergence},
the curves from all three methods converge when the number of epochs increases,
which reflects the stability of training process.
In addition, the curves of the proposed C-FCRN+Aux method are significantly lower compared to the other two for all four datasets,
which demonstrate that the proposed method allows to train a model that yields better model-fitting with the deep supervisions from the AuxCNNs.
This analysis of validation loss over the training process is consistent with the results shown in Tables~\ref{tab:mae} and Figure~\ref{fig:mre},
and reflects the better model fitting and generalization of our DRM to the validation data.

\subsection{Computation efficiency comparison}

The computation efficiencies of the seven deep convolutional neural network-based methods, including the proposed method, FCRN, C-FCRN-Only, Count-Ception, StructRegNet, U-Net and Mask R-CNN, were compared.
The average processing time on testing images with the same GPU settings was employed as the comparison metric.
Table~\ref{tab:time} shows that the proposed method costs comparable counting time compared to the FCRN and the C-FCRN-Only methods.
The Count-Ception method is the more time-consuming one in comparison to the other three regression-based methods.
In the Count-Ception method, max-pooling layers are not employed in the network,
and filters with large spatial size ($5 \times 5$ pixels) are employed for extracting multi-scale features from images.
These two reasons induce a large amount of convolution operations between high-dimension feature maps and large-sized filters, therefore, leading to the high computation workload in the Count-Ception method.

Density regression-based methods are less time-consuming than the three detection-based methods (StructRegNet, U-Net, and Mask R-CNN).
The main reason is that the non-maximum suppression post-processing for cell detection costs a considerable amount of time. Mask R-CNN takes particularly longer time because of its superior larger network size and its aim at predicting separate masks for each cell, which is a much more complex task compared to the cell counting task.

\vspace{-0.2in}
\begin{table}[h]
	\caption{Computational Efficiency Comparison}
	\vspace{-0.1in}
	\footnotesize{
	\begin{center}
		\begin{threeparttable}
		\begin{tabular}{c c c c c c}
			\hline\hline\noalign{\smallskip}
			\textbf{Seconds/image}		& \textbf{Bacterial} & \textbf{Bone marrow}& \textbf{Colorectal} & \textbf{hESC} \\
			\textbf{}						& \textbf{cells } 	    	& \textbf{cells}& \textbf{cancer cells} & \textbf{} \\
			Mask R-CNN	& $0.55279$ 	& $0.89527$ & --- 	& ---\\
			U-Net & $0.07646$	& $0.16125$ 	&--- & ---\\
			StructRegNet	&  $0.06648 $	& $0.08035 $ 	& $0.18690$ & $0.36167$\\
			FCRN & $0.00468$	& $0.02568$ 	&$0.017$ & $0.01901$\\
			C-FCRN-Only	& $0.00511$ 	& $0.02846$ & $0.01925 $ 	& $0.02134$\\
			Count-Ception	&  $0.25111 $	& $0.18185 $ 	& $0.16308$ & $0.19208$\\
			C-FCRN+Aux &  $0.00554$	& $0.03113$ 	& $0.02233$	& $0.02275$\\
			\hline\hline
		\end{tabular}
		\begin{tablenotes}
		\item Seconds/image represents the processing time for one image.
		\end{tablenotes}
		\end{threeparttable}
	\end{center}
	}
	\label{tab:time}
	\vspace{-0.2in}
\end{table}

\section{Discussion}
\label{sec:discussion}

The method proposed in this study combines the advantage of FCRN design with concatenation layers and a deeply-supervised learning strategy.
It solves the two shortcomings that exist in the original FCRN.
The concatenated layers integrates multi-scale features extracted from non-adjacent layers
to improve the granularity of the extracted features and further support the density map estimation.
The deeply-supervised learning strategy permits a direct supervision from AuxCNNs on learning its intermediate layers to mitigate the potential varnishing gradient issue and improve the cell counting performance.
The results on four image datasets show superior cell counting performance of the proposed method compared to the other eight state-of-the-art methods.
In addition, compared to the original FCRN, the proposed method improve the counting performance on four datasets ranging from $13\%$ to $31\%$ in terms of MAE.
The computational efficiency of the proposed method is comparable to other density regression-based methods.
The proposed method is capable of processing arbitrary-size images and estimating their density maps
by use of fully convolutional layers in the C-FCRN. The proposed method could also be applied to heterogeneous cell assemblies, if cell types of interest are annotated in the training images. This deeply supervised learning framework will encourage the trained DRM to focus on the cell types of interest but consider cells of other types as background.

The proposed method, other four regression-based and four detection-based methods were investigated on four challenging datasets.
In general, the density regression-based methods yielded better performance and had three advantages over the detection-based methods.
First, the regression-based methods count cells without cell detection, which can avoid challenging cell detection issues that commonly exist in microscopy images.
Second, density regression-based methods are convenient for deployment, since they do not require trivial post-processings such as thresholding and non-maximum suppression.
Thirdly, density regression-based methods can count cells more efficiently, i.e. the counting for an image of $512\times512$ pixels takes about $20 ms$.
The three advantages enable the density-regression based methods to be potentially applied to real-time clinical applications.
In addition, it should be noted that even though the detection-based methods yield lower performance on this cell counting task, they are more suitable for the segmentation of cells of other types for other applications~\cite{zhang2018panoptic,johnson2018adapting}. Generally, for those cell types of interest, the cells in the acquired microscopy images are less overlapped and the cell masks can be fully annotated.
\textcolor{black}{
In addition, the kernel sizes shown in Eq.~\ref{eq:kernel} is determined by $K_G$, which is chosen according to the sizes of cells in the processed image to guarantee that the touching areas between occluded cells have been appropriately represented on the related density map.
In this study, the radii of cells in the four datasets are less than $8$ pixels. We then set the kernel size $(2K_G+1)\times (2K_G+1)$ to $21\times21$ pixels.} 

In the current study, all images are pre-processed by simply normalizing the intensities to the range of $[0,1]$ to increase the stability of the model training process.
In the future, we will investigate other image denoising and/or image enhancement methods
to more accurately count cells for images that exhibit highly inhomogeneous tissue backgrounds and noises, or yield low image contrast.
Also, the cell centroids used for generating ground truth density maps in the three experimental datasets were
manually annotated by human experts, which may be subject to subjective errors.
This might be one of the reasons that the MREs of these three experimental datasets (shown in Figure~\ref{fig:mre}) were higher than that of the synthetic bacterial dataset.
More accurate annotation strategies will be investigated to reduce the uncertainty in generating ground truth density maps.
In this study, a uniform network architecture of C-FCRN+Aux was applied to learn DRMs separately on each of the four distinct datasets.
We will adapt some other variants of FCRNs in the future that aim at crowd counting tasks~\cite{zhang2015cross,walach2016learning,sindagi2017cnn} for varied datasets.

\section{Conclusion}
\label{sec:conclusion}

A deeply-supervised density regression model is proposed in this study for accurately and robustly counting the number of cells in microscopy images.
The proposed method is capable of processing varied-size images containing dense cell clusters,
large variations of cell morphology and inhomogeneous background noise.
Extensive experiments based on four datasets representing different image modalities and image acquisition techniques demonstrated the efficiency, robustness, and generality of the proposed method. The proposed method can be potentially to be applied to real-time clinical applications. It also holds the promise to be applied to a number of different problems, such as object counting (other than cells) in crowded scenes.

\section*{Acknowledgment}
This work was supported in part by award NIH R01EB020604, R01EB023045, R01NS102213, R01CA233873, R21CA223799, and a grant from Children Discovery Institute (LSK).
The dataset of human embryonic stem cells are provided by Solnica-Krezel group at Washington University School of Medicine.
The authors greatly appreciate the useful discussion with Dr. Su Ruan at The University at Rouen and Dr. Frank Brooks at The University of Illinois at Urbana-Champaign. The authors would like to thank the anonymous reviewers for valuable comments and suggestions.
\linenumbers

\section*{References}

\bibliography{references}

\end{document}